\documentclass[12pt,english,12pt,aps,prd,manuscript]{revtex4}
\usepackage[T1]{fontenc}
\usepackage[latin9]{inputenc}
\usepackage{color}
\usepackage{babel}
\usepackage{textcomp}
\usepackage{amstext}
\usepackage{graphicx}
\usepackage{esint}
\usepackage[unicode=true,pdfusetitle,
 bookmarks=true,bookmarksnumbered=false,bookmarksopen=false,
 breaklinks=false,pdfborder={0 0 0},pdfborderstyle={},backref=false,colorlinks=true]
 {hyperref}
\hypersetup{
 linkcolor=blue, citecolor=cyan}

\makeatletter
\@ifundefined{textcolor}{}
{%
 \definecolor{BLACK}{gray}{0}
 \definecolor{WHITE}{gray}{1}
 \definecolor{RED}{rgb}{1,0,0}
 \definecolor{GREEN}{rgb}{0,1,0}
 \definecolor{BLUE}{rgb}{0,0,1}
 \definecolor{CYAN}{cmyk}{1,0,0,0}
 \definecolor{MAGENTA}{cmyk}{0,1,0,0}
 \definecolor{YELLOW}{cmyk}{0,0,1,0}
}

\usepackage{xcolor}

\makeatother

\begin{document}
\title{Anomalous magnetohydrodynamics with longitudinal boost invariance
and chiral magnetic effect}
\author{Irfan Siddique}
\affiliation{Department of Modern Physics, University of Science and Technology
of China, Hefei 230026, China}
\author{Ren-jie Wang}
\affiliation{Department of Modern Physics, University of Science and Technology
of China, Hefei 230026, China}
\author{Shi Pu}
\affiliation{Department of Modern Physics, University of Science and Technology
of China, Hefei 230026, China}
\author{Qun Wang}
\affiliation{Department of Modern Physics, University of Science and Technology
of China, Hefei 230026, China}
\begin{abstract}
We study relativistic magnetohydrodynamics with longitudinal boost
invariance in the presence of chiral magnetic effects and finite electric
conductivity. With initial magnetic fields parallel or anti-parallel
to electric fields, we derive the analytic solutions of electromagnetic
fields and the chiral number and energy density in an expansion of
several parameters determined by initial conditions. The numerical
solutions show that such analytic solutions work well in weak fields
or large chiral fluctuations. We also discuss the properties of electromagnetic
fields in the laboratory frame.
\end{abstract}
\maketitle

\section{Introduction}


Recently some novel transport phenomena of chiral (massless) fermions
in strong electromagnetic (EM) fields have been extensively studied
in relativistic heavy ion collisions and condensed matter physics.
One of them is the chiral magnetic effect (CME): an electric current
can be induced by the strong magnetic field when the numbers of left
and right handed fermions are not equal \citep{Vilenkin1980a,Kharzeev:2007jp,Fukushima:2008xe}.
Similarly the strong magnetic field can also lead to the chiral separation
effect (CSE) for the chiral charge current. These effects are associated
with the chiral anomaly and can be described by chiral kinetic equations
(CKE). The CKE are derived from various approaches, e.g. the path
integral \citep{Stephanov:2012ki,Chen:2013iga,Chen:2014cla}, the
Hamiltonian approach \citep{Son:2012wh,Son:2012zy}, the quantum kinetic
theory via Wigner functions \citep{Gao:2012ix,Chen:2012ca,Gao:2015zka,Hidaka:2016yjf,Hidaka:2017auj,Gao:2017gfq,Hidaka:2018mel,Gao:2018wmr,Huang:2018wdl},
and the world-line formalism \citep{Mueller:2017arw,Mueller:2017lzw}.
The chiral separation can also be induced by an electric field, which
is called the chiral electric separation effect (CESE) \citep{Huang:2013iia,Pu:2014cwa,Jiang:2014ura,Pu:2014fva}.
If the electric field is perpendicular to the magnetic field, a Hall
current for chiral fermions is expected, which is called chiral Hall
separation effect (CHSE) \citep{Pu:2014fva}. The chiral particle
production in strong EM fields are found to be directly connected
to the Schwinger mechanism \citep{Fukushima:2010vw,Warringa:2012bq},
and similar calculation has been done analytically via the world-line
formalism \citep{Copinger:2018ftr} and Wigner functions \citep{Sheng:2018jwf}.
Recent reviews about chiral transport phenomena can be found in Ref.
\citep{Bzdak:2012ia,Fukushima:2012vr,Kharzeev:2013ffa,Kharzeev:2015kna}. 


The chiral transport phenomena are expected to have observables in
relativistic heavy ion collisions in which very strong magnetic fields
of the order $B\sim10^{18}\,\mathrm{G}$ are produced \citep{Bzdak:2011yy,Deng:2012pc,Roy:2015coa,Li:2016tel}.
At the very early stage of the quark-gluon plasma (QGP), the topological
fluctuations in non-Abelian gauge fields give rise to the imbalance
of chirality from event to event (event-by-event). Such an imbalance
of chirality may lead to the charge separation with respect to the
reaction plane in heavy ion collisions. The STAR collaboration have
observed the charge separation in Au+Au collisions \citep{Abelev2009,Abelev2010}.
However, due to the huge backgrounds from collective flows \citep{Khachatryan:2016got,Sirunyan:2017quh}
it is a challenge to extract the weak CME signal from the overwhelming
backgroud. It is expected that the ongoing isobar collision experiment
at STAR may shed light on the CME signal (see e.g. Ref. \citep{Skokov:2016yrj}
for discussions on isobar collisions). 


In order to extract the CME signal, we need the precise simulation
of the QGP evolution in the time-evolving EM field. One approach is
through the simulation of the CKE. Very recently, the boost invariant
formulation of the CKE has been done with the chiral circular displacement
introduced \citep{Ebihara:2017suq}. The CKE has been solved numerically
in heavy ion collisions \citep{Sun:2016nig,Huang:2017tsq}. Another
approach is the classical statistical simulation based on solving
the coupled equations of Yang-Mills and Dirac applied to heavy ion
collisions \citep{Mace:2016svc,Mace:2016shq,Berges:2017igc}. Besides
the relativistic hydrodynamic is a widely-used model in relativistic
heavy ion collisions.

The relativistic hydrodynamic model is one of the main approaches
to the QGP evolution \citep{Romatschke2007,Luzum2008,Song2008b,Song:2008si,Schenke:2011bn,Roy:2012jb,Niemi:2012ry}.
A natural extension of the hydrodynamic model in the presence of the
magnetic field is the magento-hydrodynamics (MHD), which is hydrodynamics
coupled with Maxwell's equations. The ideal MHD equations with longitudinal
boost invariance and a transverse magnetic field has been calculated
\citep{Pu:2016ayh,Roy:2015kma}, where the magnetic field decays as
$\sim1/\tau$ with $\tau$ being the proper time, much slower than
in vacuum \citep{Kharzeev:2007jp}. The magnetization effect has also
been systematically studied \citep{Pu:2016ayh}. Later the calculation
has been extended to 2+1 dimensions \citep{Pu:2016bxy,Pu:2016rdq}.
There is an enhancement of the elliptic flow $v_{2}$ of $\pi^{-}$
from the external magnetic field \citep{Roy:2017yvg}. Recently the
MHD with the longitudinal boost invariance has been extended to include
the finite conductivity in the Gubser flow \citep{Shokri:2018qcu}.
Readers may look at Ref. \citep{Inghirami:2016iru} for recent numerical
simulations of the ideal MHD. 



In this work, we will consider the relativistic MHD in the presence
of the CME and finite conductivity. Usually the numerical simulationsof
MHD with the CME could be very unstable because of chirality instability
\citep{Akamatsu2013}. Therefore stable analytic solutions in some
special cases are very important for providing a test of numerical
simulations and a simple physical picture for such a complicated process.
As a first attempt, we will consider the MHD with the longitudinal
boost invariance. To avoid the acceleration of the fluid by the EM
field, we will assume an electric charge neutral fluid. We then search
for the EM fields that can keep the Bjorken fluid velocity unchanged.
It is very similar to the case of the force-free magnetic field discussed
in classical electrodynamics \citep{Chandrasekhar285,Woltjer489}.
To solve the coupled equations of the anomalous conservation equation
and Maxwell's equations, we assume that the terms proportional to
the anomaly constant (proportional to the Planck constant $\hbar$)
are perturbations, this is equivalent to an expansion in $\hbar$.
We will compare our approximate analytic solutions with the numerical
results. Finally we compute the EM field in the laboratory frame and
discuss the coupling between the EM field and the chiral current.


The organization of the paper is as follows. In Sec. \ref{sec:MHD-with-CME},
we give a brief review for the relativistic MHD with the CME. In Sec.
\ref{sec:Longitudinal-boost-invariant}, we assume the form of the
fluid velocity in longitudinal boost invariance. We choose a configuration
of the EM field that is orthogonal to the fluid velocity. In Sec.
\ref{subsec:For-EoS-1} and \ref{subsec:For-EoS-2}, we solve Maxwell's
equations coupled with the anomalous conservation equation for the
chiral charge. We obtain the approximate analytic solutions for two
different equations of state. We compare our approximate analytic
solutions with numerical ones. In Sec. \ref{subsec:Discussion-and-physical},
we compute the EM field in the laboratory frame to show the consistence
with previous results. Finally we make a summary of our results in
Sec. \ref{sec:Summary-and-conclusion}. 

Throughout this work, we will use the metric $g_{\mu\nu}=\mathrm{diag}\{+,-,-,-\}$,
thus, the fluid velocity satisfies $u^{\mu}u_{\mu}=1$, and the orthogonal
projector to the fluid four-velocity is $\Delta^{\mu\nu}=g^{\mu\nu}-u^{\mu}u^{\nu}$.
We also choose Levi-Civita tensor satisfying $\epsilon^{0123}=-\epsilon_{0123}=+1$
and $\epsilon^{\mu\nu\alpha\beta}\epsilon_{\mu\nu\rho\sigma}=-2!(g_{\rho}^{\alpha}g_{\sigma}^{\beta}-g_{\sigma}^{\alpha}g_{\rho}^{\beta})$.

\section{Anomalous magnetohydrodynamics}

\label{sec:MHD-with-CME}In this section, we will give a brief preview
to the relativistic MHD with CME which is called anomalous magnetohydrodynamics.
The MHD equations consist of conservation equations and Maxwell's
equations (see, e.g., Ref. \citep{Caldarelli:2008ze,Gedalin:PRE1995,Huang:2009ue,Roy:2015kma,Pu:2016ayh,Pu:2016bxy,Pu:2016rdq}
for details). The energy-momentum conservation equation reads 
\begin{equation}
\partial_{\mu}T^{\mu\nu}=0,\label{eq:EMT_01}
\end{equation}
where $T^{\mu\nu}$is the energy momentum tensor including the contributions
from the fluid and the EM fields 
\begin{equation}
T^{\mu\nu}=T_{F}^{\mu\nu}+T_{EM}^{\mu\nu}.\label{eq:total-t}
\end{equation}
The fluid part has the usual form
\begin{equation}
T_{F}^{\mu\nu}=\varepsilon u^{\mu}u^{\nu}\text{\textminus}(p+\Pi)\Delta^{\mu\nu}+\pi^{\mu\nu},
\end{equation}
where $\varepsilon$ and $p$ are the energy density and pressure
respectively, $u^{\mu}=\gamma(1,\mathbf{v})$ is the fluid velocity
satisfying $u^{\mu}u_{\mu}=1$, $\Delta^{\mu\nu}=g^{\mu\nu}-u^{\mu}u^{\nu}$
is the projector, and $\Pi$ and $\pi^{\mu\nu}$ are bulk viscous
pressure and shear viscous tensor respectively. For simplicity, we
will neglect viscous effects in this paper, i.e. $\Pi=\pi^{\mu\nu}=0$.
The EM field part of the energy-momentum tensor reads 
\begin{equation}
T_{EM}^{\mu\nu}=-F^{\mu\lambda}F_{\;\lambda}^{\nu}+\frac{1}{4}g^{\mu\nu}F^{\rho\sigma}F_{\rho\sigma}.\label{eq:EM_tensor_01}
\end{equation}
One can introduce the four-vector form of the electric and magnetic
fields in terms of the fluid velocity 
\begin{equation}
E^{\mu}=F^{\mu\nu}u_{\nu},\;B^{\mu}=\frac{1}{2}\epsilon^{\mu\nu\alpha\beta}u_{\nu}F_{\alpha\beta},\label{eq:EB_def01}
\end{equation}
which satisfy $u^{\mu}E_{\mu}=0$ and $u^{\mu}B_{\mu}=0$ meaning
that both $E^{\mu}$ and $B^{\mu}$ are space-like. Then, the EM field
strength tensor can be put into the form 
\begin{equation}
F^{\mu\nu}=E^{\mu}u^{\nu}-E^{\nu}u^{\mu}+\epsilon^{\mu\nu\alpha\beta}u_{\alpha}B_{\beta},\label{eq:F_01}
\end{equation}
Inserting the above formula into Eq. (\ref{eq:EM_tensor_01}), we
obtain the complete form of the energy-momentum tensor from Eq. (\ref{eq:total-t})
\begin{eqnarray}
T^{\mu\nu} & = & (\varepsilon+p+E^{2}+B^{2})u^{\mu}u^{\nu}-(p+\frac{1}{2}E^{2}+\frac{1}{2}B^{2})g^{\mu\nu}\nonumber \\
 &  & -E^{\mu}E^{\nu}-B^{\mu}B^{\nu}-u^{\mu}\epsilon^{\nu\lambda\alpha\beta}E_{\lambda}B_{\alpha}u_{\beta}-u^{\nu}\epsilon^{\mu\lambda\alpha\beta}E_{\lambda}B_{\alpha}u_{\beta},\label{eq:totEMtensor_01}
\end{eqnarray}
where $E$ and $B$ are defined by 
\begin{equation}
E^{\mu}E_{\mu}=\text{\textminus}E^{2},\;B^{\mu}B_{\mu}=\text{\textminus}B^{2}.
\end{equation}

The conservations equations are 
\begin{eqnarray}
\partial_{\mu}j_{e}^{\mu} & = & 0,\nonumber \\
\partial_{\mu}j_{5}^{\mu} & = & -e^{2}CE\cdot B,\label{eq:currents_con_01}
\end{eqnarray}
where $j_{e}^{\mu}$ is the electric charge current and $j_{5}^{\mu}$
is the chiral (axial) charge current. Note that the chiral anomaly
term appears in the second line of Eq. (\ref{eq:currents_con_01})
with $C=1/(2\pi^{2})$. These currents can be decomposed into three
parts 

\begin{eqnarray}
j_{e}^{\mu} & = & n_{e}u^{\mu}+\sigma E^{\mu}+\xi B^{\mu},\nonumber \\
j_{5}^{\mu} & = & n_{5}u^{\mu}+\sigma_{5}E^{\mu}+\xi_{5}B^{\mu},\label{eq:current_02}
\end{eqnarray}
where $n_{e}$ and $n_{5}$ are the electric and chiral charge density
respectively, $\sigma$ and $\sigma_{5}$ are the electric and chiral
electric conductivity respectively \citep{Huang:2013iia,Pu:2014cwa,Pu:2014fva},
and $\xi$ and $\xi_{5}$ are associated with the CME and CESE \citep{Fukushima:2008xe,Gao:2012ix,Chen:2012ca}
which are given by 
\begin{equation}
\xi=eC\mu_{5},\;\xi_{5}=eC\mu_{e}.\label{eq:CME_coefficient_01}
\end{equation}
For simplicity, we will neglect all other dissipative effects in $j_{e}^{\mu}$
and $j_{5}^{\mu}$ such as the heat conducting flow. The chiral electric
conductivity $\sigma_{5}$ is usually parametrized as $\sigma_{5}\propto\mu_{e}\mu_{5}$
in the small $\mu_{e}$ and $\mu_{5}$ limit \citep{Huang:2013iia,Pu:2014cwa,Pu:2014fva}. 

Maxwell's equations can be put into the following form 
\begin{eqnarray}
\partial_{\mu}F^{\mu\nu} & = & j_{e}^{\nu},\label{eq:Maxwell_01a}\\
\partial_{\mu}(\epsilon^{\mu\nu\alpha\beta}F_{\alpha\beta}) & = & 0.\label{eq:Maxwell_01b}
\end{eqnarray}

To close the system of equations, we need to choose the equations
of state (EoS) for the thermodynamic quantities. In the dense limit
with high chemical potentials, we use 
\begin{eqnarray}
\varepsilon & = & c_{s}^{-2}p,\nonumber \\
n_{e} & = & a\mu_{e}(\mu_{e}^{2}+3\mu_{5}^{2}),\nonumber \\
n_{5} & = & a\mu_{5}(\mu_{5}^{2}+3\mu_{e}^{2}),\label{eq:eos_01}
\end{eqnarray}
where $a$ is a dimensionless constant and $c_{s}$ is the speed of
sound also taken as a constant. On the other hand, in the hot limit
with high temperatures, we use 
\begin{eqnarray}
\varepsilon & = & c_{s}^{-2}p,\nonumber \\
n_{e} & = & a\mu_{e}T^{2},\nonumber \\
n_{5} & = & a\mu_{5}T^{2},\label{eq:eos_02}
\end{eqnarray}
where $a$ is again a dimensionless constant. Note that the value
of $a$ in Eq. (\ref{eq:eos_02}) is different from that in Eq. (\ref{eq:eos_01}).
For the ideal fluid, we have $a=1/(3\pi^{2})$ and $a=1/3$ for Eq.
(\ref{eq:eos_01}) and (\ref{eq:eos_02}) respectively \citep{Pu:2011vr,Gao:2012ix}. 

Usually the electric field would accelerate charged particles and
the charged fluid. To avoid such a problem, we simply set the chemical
potential for electric charge vanishing, $\mu_{e}=0$, which also
leads to $n_{e}=\sigma_{5}=\xi_{5}=0$. Such a condition means the
fluid is neutral: the number of positively charged particles is the
same as that of negatively charged particles. Actually we look for
a special configuration of EM fields coupled with the media, very
similar to the force-free case in classical electrodynamics. In Sec.
\ref{subsec:Discussion-and-physical}, we will discuss the details
and check the consistence of this assumption. 

Here are the whole system of equations we are going to solve: conservation
equations (\ref{eq:EMT_01}, \ref{eq:currents_con_01}), Maxwell's
equations (\ref{eq:Maxwell_01a}, \ref{eq:Maxwell_01b}), constitutive
equations (\ref{eq:totEMtensor_01}, \ref{eq:F_01}, \ref{eq:current_02}),
and equations of state (\ref{eq:eos_01},\ref{eq:eos_02}). 

\section{Equations with longitudinal boost invariance}

\label{sec:Longitudinal-boost-invariant}We assume that the fluid
has longitudinal boost invariance. It is convenient to introduce the
Milne coordinates $z=\tau\sinh\eta$ and $t=\tau\cosh\eta$, with
$\tau=\left(t^{2}-z^{2}\right)^{1/2}$ being the proper time and $\eta=\frac{1}{2}\ln[(t+z)/(t-z)]$
being the space-time rapidity. The fluid velocity with longitudinal
boost invariance can be written as \citep{Bjorken:1982qr},
\begin{equation}
u^{\mu}=\left(\cosh\eta,0,0,\sinh\eta\right)=\gamma(1,0,0,z/t),\label{eq:Bjokren_velocity_01}
\end{equation}
where $\gamma=\cosh\eta$ is the Lorentz contraction factor. 

For simplicity we neglect the EM field in the longitudinal direction,
so the general form of the EM field satisfying $u\cdot E=u\cdot B=0$
is 

\begin{eqnarray}
E^{\mu} & = & \left(0,E\text{\ensuremath{\cos}}\zeta,E\sin\zeta,0\right),\nonumber \\
B^{\mu} & = & \left(0,B\text{\ensuremath{\cos}}\varphi,B\sin\varphi,0\right),\label{eq:em-lbi}
\end{eqnarray}
where $\zeta$ and $\varphi$ are the azimuthal angle of the electric
and magnetic field in the transverse plane respectively. To search
for possible analytic solutions, we assume that $E^{\mu}$ and $B^{\mu}$
will always be in the transverse plane and that $E,B,\zeta,\varphi$
are only functions of $\tau$. We can further simplify the probelm
by assuming that $E^{\mu}$ and $B^{\mu}$ are parallel or anti-parallel.
Without loss of generality, the EM field can be put in the $y$ direction
\begin{equation}
E^{\mu}=(0,0,\chi E(\tau),0),\;B^{\mu}=(0,0,B(\tau),0),\label{eq:EB_02}
\end{equation}
where $\chi=\pm1$. We will check the self-consistence of these assumptions
after we find the solution in Sec. \ref{subsec:Discussion-and-physical}.
Note that the authors of Ref. \citep{Shokri:2017xxn} have found another
possible configuration of the EM fields in the absence of the chiral
magnetic effect, in which the direction of the electric and magnetic
field depends on $\eta$. As this configuration is irrelevant to the
heavy ion collisions, we will not consider it in this paper. 

By projecting the energy-momentum conservation equation (\ref{eq:EMT_01})
onto the spatial direction, $\Delta_{\mu\alpha}\partial_{\nu}T^{\mu\nu}=0$,
we obtain the acceleration of the fluid velocity 
\begin{eqnarray}
(u\cdot\partial)u_{\alpha} & = & \frac{1}{(\varepsilon+p+E^{2}+B^{2})}[\Delta_{\mu}^{\nu}\partial_{\nu}(p+\frac{1}{2}E^{2}+\frac{1}{2}B^{2})+\Delta_{\mu\alpha}(E\cdot\partial)E^{\mu}+E_{\alpha}(\partial\cdot E)\nonumber \\
 &  & +\Delta_{\mu\alpha}(B\cdot\partial)B^{\mu}+B_{\alpha}(\partial\cdot B)+\epsilon^{\nu\lambda\rho\sigma}E_{\lambda}B_{\rho}u_{\sigma}(\partial_{\nu}u_{\alpha})+(\partial\cdot u)\epsilon_{\alpha\lambda\rho\sigma}E^{\lambda}B^{\rho}u^{\sigma}\nonumber \\
 &  & +\Delta_{\mu\alpha}(u\cdot\partial)\epsilon^{\mu\lambda\rho\sigma}E_{\lambda}B_{\rho}u_{\sigma}].\label{eq:Du_01}
\end{eqnarray}
According to our assumption that the electric and magnetic field are
constant in transverse coordinates $(x,y)$, we have $(E\cdot\partial)E^{\mu}=(\partial\cdot E)=(B\cdot\partial)B^{\mu}=(\partial\cdot B)=0$.
Also, if $p$, $E^{\mu}$ and $B^{\mu}$ are only the functions of
$\tau$, the first term inside the square brackets are vanishing.
So we obtain the non-acceleration of the fluid velocity 
\begin{equation}
(u\cdot\partial)u_{\alpha}=0,\label{eq:non-acc}
\end{equation}
which means that the fluid velocity always takes the value in Eq.
(\ref{eq:Bjokren_velocity_01}). This is consistent to the previous
assumption that the fluid is charge neutral. 

The energy conservation equation can be obtained by a contraction
of $u_{\mu}$ with Eq. (\ref{eq:EMT_01}) or $u_{\mu}\partial_{\nu}T^{\mu\nu}=0$,
\begin{eqnarray}
 &  & (u\cdot\partial)(\varepsilon+\frac{1}{2}E^{2}+\frac{1}{2}B^{2})+(\varepsilon+p+E^{2}+B^{2})(\partial\cdot u)\nonumber \\
 & = & u_{\mu}(E\cdot\partial)E^{\mu}+u_{\mu}(B\cdot\partial)B^{\mu}+\epsilon^{\nu\lambda\alpha\beta}\partial_{\nu}(E_{\lambda}B_{\alpha}u_{\beta})\nonumber \\
 &  & +u_{\mu}(u\cdot\partial)\epsilon^{\mu\lambda\alpha\beta}E_{\lambda}B_{\alpha}u_{\beta}.
\end{eqnarray}
With Eq. (\ref{eq:EB_02}), the above equation is reduced to 
\begin{equation}
(u\cdot\partial)(\varepsilon+\frac{1}{2}E^{2}+\frac{1}{2}B^{2})+(\varepsilon+p+E^{2}+B^{2})(\partial\cdot u)=0.\label{eq:energy_density_02}
\end{equation}

Now we look at Maxwell's equations. Inserting Eq. (\ref{eq:EB_02})
for the EM fields into Eq. (\ref{eq:Maxwell_01a}) yields for $\nu=y$
\begin{equation}
\frac{d}{d\tau}E+\frac{1}{\tau}E+\sigma E+\chi\xi B=0,\label{eq:Maxwell_02a}
\end{equation}
where we have used $d/d\tau\equiv(u\cdot\partial)$. For other indices
$\nu=t,x,z$, we obtain identities using $\mu_{e}=0$ and $n_{e}=0$.
Similarly, from Eq. (\ref{eq:Maxwell_01b}), we obtain for $\nu=y$
\begin{equation}
\frac{d}{d\tau}B+\frac{B}{\tau}=0.\label{eq:Maxwell_02b}
\end{equation}
For other indices $\nu=t,x,z$, we obtain identities using $\mu_{e}=0$
and $n_{e}=0$. 

Using the simplified Maxwell's equations (\ref{eq:Maxwell_02a}) and
(\ref{eq:Maxwell_02b}), we can rewrite Eq. (\ref{eq:energy_density_02})
into a compact form
\begin{equation}
\frac{d}{d\tau}\varepsilon+(\varepsilon+p)\frac{1}{\tau}-\sigma E^{2}-\chi\xi EB=0.\label{eq:energy_density_03}
\end{equation}
This equation can also be derived by rewritten Eq. (\ref{eq:EMT_01})
as 
\begin{equation}
\partial_{\mu}T_{F}^{\mu\nu}=-\partial_{\mu}T_{EM}^{\mu\nu}=F^{\nu\lambda}j_{e\lambda},\label{eq:EM_con_02}
\end{equation}
Contracting the above equation with $u_{\nu}$ yields $u_{\nu}\partial_{\mu}T_{F}^{\mu\nu}=-E^{\lambda}j_{e\lambda}$,
which is consistent with Eq. (\ref{eq:energy_density_03}).

From Eq. (\ref{eq:currents_con_01}) and using $\mu_{e}=0$, the (anomalous)
conservation equation of the chiral charge can be reduced to 
\begin{eqnarray}
\frac{d}{d\tau}n_{5}+\frac{n_{5}}{\tau} & = & e^{2}C\chi EB.\label{eq:conserved_current_02}
\end{eqnarray}
The conservation equation for $j_{e}^{\mu}$ is automatically satisfied
with $\mu_{e}=0$ and $E^{\mu},B^{\mu}$ taking the form of Eq. (\ref{eq:EB_02}). 

Before we end this section, we make some remarks about the simplified
equations with longitudinal boost invariance. To enforce the fluid
velocity not accelerated, the EM field are assumed to take the form
as Eq. (\ref{eq:EB_02}). Using Maxwell's equations the energy conservation
equation $u_{\mu}\partial_{\nu}T^{\mu\nu}=0$ is reduced to Eq. (\ref{eq:energy_density_03}).
The momentum conservation equation $\Delta_{\mu\alpha}\partial_{\nu}T^{\mu\nu}=0$
is reduced to Eq. (\ref{eq:non-acc}) meaning that the fluid velocity
always takes value in (\ref{eq:Bjokren_velocity_01}). Maxwell's equations
(\ref{eq:Maxwell_01a}, \ref{eq:Maxwell_01b}) are simplified to Eqs.
(\ref{eq:Maxwell_02a}, \ref{eq:Maxwell_02b}). The chiral charge
conservation equation in Eq. (\ref{eq:currents_con_01}) is simplified
to Eq. (\ref{eq:conserved_current_02}). 

\section{Analytic solutions}

\label{sec:Approximate-analytic-solutions}We will use the non-conserved
charges method \citep{Csorgo:2003rt,Shokri:2017xxn} to solve Eqs.
(\ref{eq:Maxwell_02a}, \ref{eq:Maxwell_02b}, \ref{eq:energy_density_03},
\ref{eq:conserved_current_02}) with the EoS (\ref{eq:eos_01}) or
(\ref{eq:eos_02}). 

The non-conserved charges method is to solve the equation for $f(\tau)$
in the following form 
\begin{equation}
\frac{d}{d\tau}f(\tau)+m\frac{f(\tau)}{\tau}=f(\tau)\frac{d}{d\tau}\lambda(\tau),\label{eq:noncharge_01}
\end{equation}
where $m$ is a constant and $\lambda(\tau)$ is a known function.
The general solution is
\begin{equation}
f(\tau)=f(\tau_{0})\exp\left[\lambda(\tau)-\lambda(\tau_{0})\right]\left(\frac{\tau_{0}}{\tau}\right)^{m},\label{eq:general_form_01}
\end{equation}
where $\tau_{0}$ is an initial proper time and $f(\tau_{0})$ is
determined by an initial value at $\tau_{0}$. In this paper we will
rewrite Eqs. (\ref{eq:Maxwell_02a}, \ref{eq:Maxwell_02b}, \ref{eq:energy_density_03},
\ref{eq:conserved_current_02}) into the form of Eq. (\ref{eq:noncharge_01})
and obtain the solutions in the form of Eq. (\ref{eq:general_form_01}). 

Note that generally $f$ can also be a function of rapidity $\eta$
\citep{Csorgo:2003rt,Shokri:2017xxn}. However, in this paper we focus
on the central rapidity region in heavy ion collisions which implies
$\eta\simeq0$ with longitudinal boost invariance, therefore we will
not consider the rapidity dependence. 

From Eq. (\ref{eq:Maxwell_02b}), we immediately obtain
\begin{eqnarray}
B(\tau) & = & B_{0}\frac{\tau_{0}}{\tau},\label{eq:B_03}
\end{eqnarray}
where $B_{0}=B(\tau_{0})$ is the initial value of the magnetic field.
We see that the proper time behavior of the magnetic field seems to
be the same as the case without CME \citep{Pu:2016ayh,Roy:2015kma,Pu:2016bxy}.
But we will show in Sec. \ref{subsec:Discussion-and-physical} the
contribution from the CME and finite conductivity to the EM field
appear in the Lab frame.

\subsection{EoS (\ref{eq:eos_01}) }

\label{subsec:For-EoS-1}For the EoS (\ref{eq:eos_01}), we will solve
Eq. (\ref{eq:Maxwell_02a}) with Eq. (\ref{eq:conserved_current_02})
to obtain $n_{5}(\tau)$ and $E(\tau)$. Then we insert $n_{5}(\tau)$
and $E(\tau)$ into Eq. (\ref{eq:energy_density_03}) to obtain the
energy-density $\varepsilon(\tau)$. 

We need to put Eqs. (\ref{eq:Maxwell_02a}, \ref{eq:conserved_current_02})
into the form of Eq. (\ref{eq:noncharge_01}) 
\begin{eqnarray}
\frac{d}{d\tau}E+\frac{E}{\tau} & = & E\frac{d}{d\tau}\mathcal{E},\nonumber \\
\frac{d}{d\tau}n_{5}+\frac{n_{5}}{\tau} & = & n_{5}\frac{d}{d\tau}\mathcal{N},\label{eq:EN_04}
\end{eqnarray}
where 
\begin{eqnarray}
\frac{d}{d\tau}\mathcal{E} & = & -\sigma-\chi\xi\frac{B}{E},\nonumber \\
\frac{d}{d\tau}\mathcal{N} & = & \frac{e^{2}C\chi EB}{n_{5}},\label{eq:EN_03}
\end{eqnarray}
and $\xi$ is given by Eq. (\ref{eq:CME_coefficient_01}) and depends
on $n_{5}$ through the EoS (\ref{eq:eos_01}). Following Eq. (\ref{eq:general_form_01}),
the formal solutions are in the form 
\begin{eqnarray}
n_{5}(\tau) & = & n_{5,0}\exp\left[\mathcal{N}(\tau)-\mathcal{N}(\tau_{0})\right]\frac{\tau_{0}}{\tau},\nonumber \\
E(\tau) & = & E_{0}\exp\left[\mathcal{E}(\tau)-\mathcal{E}(\tau_{0})\right]\frac{\tau_{0}}{\tau},\label{eq:nE_00}
\end{eqnarray}
where $n_{5,0}=n_{5}(\tau_{0})$ and $E_{0}=E(\tau_{0})$. Inserting
the above $n_{5}(\tau)$ and $E(\tau)$ as well as $B(\tau)$ in Eq.
(\ref{eq:B_03}) into Eq. (\ref{eq:EN_03}), we obtain 
\begin{eqnarray}
\frac{d}{d\tau}x & = & -\sigma x-\frac{a_{1}}{\tau_{0}}\left(\frac{\tau_{0}}{\tau}\right)^{1/3}y^{1/3},\nonumber \\
\frac{d}{d\tau}y & = & a_{2}\frac{x}{\tau},\label{eq:xy_01}
\end{eqnarray}
where we have introduced the new variables 
\begin{eqnarray}
x(\tau) & = & \exp\left[\mathcal{E}(\tau)-\mathcal{E}(\tau_{0})\right],\nonumber \\
y(\tau) & = & \exp\left[\mathcal{N}(\tau)-\mathcal{N}(\tau_{0})\right],
\end{eqnarray}
with $x(\tau_{0})=y(\tau_{0})=1$, and $a_{1}$ and $a_{2}$ are dimensionless
constants determined by the initial conditions 
\begin{eqnarray}
a_{1} & = & eC\chi\left(\frac{n_{5,0}}{a}\right)^{1/3}\frac{B_{0}}{E_{0}}\tau_{0},\nonumber \\
a_{2} & = & \frac{e^{2}C\chi E_{0}B_{0}\tau_{0}}{n_{5,0}}.
\end{eqnarray}
Instead of solving Eqs. (\ref{eq:Maxwell_02a}, \ref{eq:conserved_current_02})
or Eq. (\ref{eq:EN_04}), now we only need to solve Eq. (\ref{eq:conserved_current_02})
with the initial condition $x(\tau_{0})=y(\tau_{0})=1$. We see that
both $a_{1}$ and $a_{2}$ are linearly proportional to the anomaly
constant $C$ which is linearly proportional to the Planck constant.
This means $a_{1}$ and $a_{2}$ are of quantum nature. 

Now we try to solve Eq. (\ref{eq:xy_01}) under some approximations.
We can rewrite Eq. (\ref{eq:xy_01}) into an integral form 
\begin{eqnarray}
x(\tau) & = & e^{-\sigma(\tau-\tau_{0})}-\frac{a_{1}}{\tau_{0}}e^{-\sigma\tau}\int_{\tau_{0}}^{\tau}d\tau^{\prime}e^{\sigma\tau^{\prime}}\left(\frac{\tau_{0}}{\tau^{\prime}}\right)^{1/3}y^{1/3}(\tau^{\prime}),\nonumber \\
y(\tau) & = & 1+a_{2}\int_{\tau_{0}}^{\tau}d\tau^{\prime}\frac{x(\tau^{\prime})}{\tau^{\prime}}.\label{eq:x-y-tau}
\end{eqnarray}
Since $a_{1}$ and $a_{2}$ terms are quantum corrections as $a_{1},a_{2}\propto\hbar$,
we can deal with these terms as perturbations to the classical terms,
so Eq. (\ref{eq:xy_01}) or (\ref{eq:x-y-tau}) can be solved order
by order in powers of $\hbar$. 

To the linear order in $\hbar$, we have the solutions for $x(\tau)$
and $y(\tau)$

\begin{eqnarray*}
x(\tau) & = & e^{-\sigma(\tau-\tau_{0})}-\frac{a_{1}}{\tau_{0}^{2/3}}e^{-\sigma\tau}[\tau_{0}^{2/3}\textrm{E}_{1/3}(-\sigma\tau_{0})-\tau^{2/3}\textrm{E}_{1/3}(-\sigma\tau)],\\
y(\tau) & = & 1+a_{2}\left[e^{\sigma\tau_{0}}-a_{1}\textrm{E}_{1/3}(-\sigma\tau_{0})\right][\textrm{E}_{1}(\sigma\tau_{0})-\textrm{E}_{1}(\sigma\tau)],
\end{eqnarray*}
where $\textrm{E}_{n}(z)\equiv\int_{1}^{\infty}dtt^{-n}e^{-zt}$ is
the generated exponential integral. Then we obtain the solutions for
$E(\tau)$ and $n_{5}(\tau)$ 
\begin{eqnarray}
E(\tau) & = & E_{0}\frac{\tau_{0}}{\tau}\left\{ e^{-\sigma(\tau-\tau_{0})}-\frac{a_{1}}{\tau_{0}^{2/3}}e^{-\sigma\tau}[\tau_{0}^{2/3}\textrm{E}_{1/3}(-\sigma\tau_{0})-\tau^{2/3}\textrm{E}_{1/3}(-\sigma\tau)]\right\} ,\nonumber \\
n_{5}(\tau) & = & n_{5,0}\frac{\tau_{0}}{\tau}\left\{ 1+a_{2}e^{\sigma\tau_{0}}[\textrm{E}_{1}(\sigma\tau_{0})-\textrm{E}_{1}(\sigma\tau)]\right\} .\label{eq:EN_A_01}
\end{eqnarray}
At early proper time, $\tau\rightarrow\tau_{0}$, we can expand $\textrm{E}_{n}(\tau)$
near $\tau_{0}$ and obtain 
\begin{eqnarray}
E(\tau) & \simeq & E_{0}\frac{\tau_{0}}{\tau}\left[e^{-\sigma(\tau-\tau_{0})}-\frac{a_{1}}{\tau_{0}}(\tau-\tau_{0})+a_{1}\frac{1+3\tau_{0}\sigma}{6\tau_{0}^{2}}(\tau-\tau_{0})^{2}\right],\nonumber \\
n_{5}(\tau) & \simeq & n_{5,0}\frac{\tau_{0}}{\tau}\left\{ 1+a_{2}\frac{\tau-\tau_{0}}{\tau_{0}}-a_{2}\frac{1+\sigma\tau_{0}}{2\tau_{0}^{2}}(\tau-\tau_{0})^{2}\right\} .
\end{eqnarray}
Finally the energy density and the pressure can be solved by using
the solutions for $\mu_{5},E,B$. From Eq. (\ref{eq:energy_density_03}),
we obtain the energy density 
\begin{eqnarray}
\varepsilon(\tau) & = & \varepsilon_{0}\left(\frac{\tau_{0}}{\tau}\right)^{1+c_{s}^{2}}(1+\Delta\varepsilon),\nonumber \\
\Delta\varepsilon(\tau) & = & \frac{1}{\varepsilon_{0}}\int_{\tau_{0}}^{\tau}d\tau^{\prime}\left(\frac{\tau^{\prime}}{\tau_{0}}\right)^{1+c_{s}^{2}}\left[\sigma E^{2}(\tau^{\prime})+\chi\xi(\tau^{\prime})E(\tau^{\prime})B(\tau^{\prime})\right].\label{eq:end_01}
\end{eqnarray}

\begin{figure}[tp]
\includegraphics[scale=0.35]{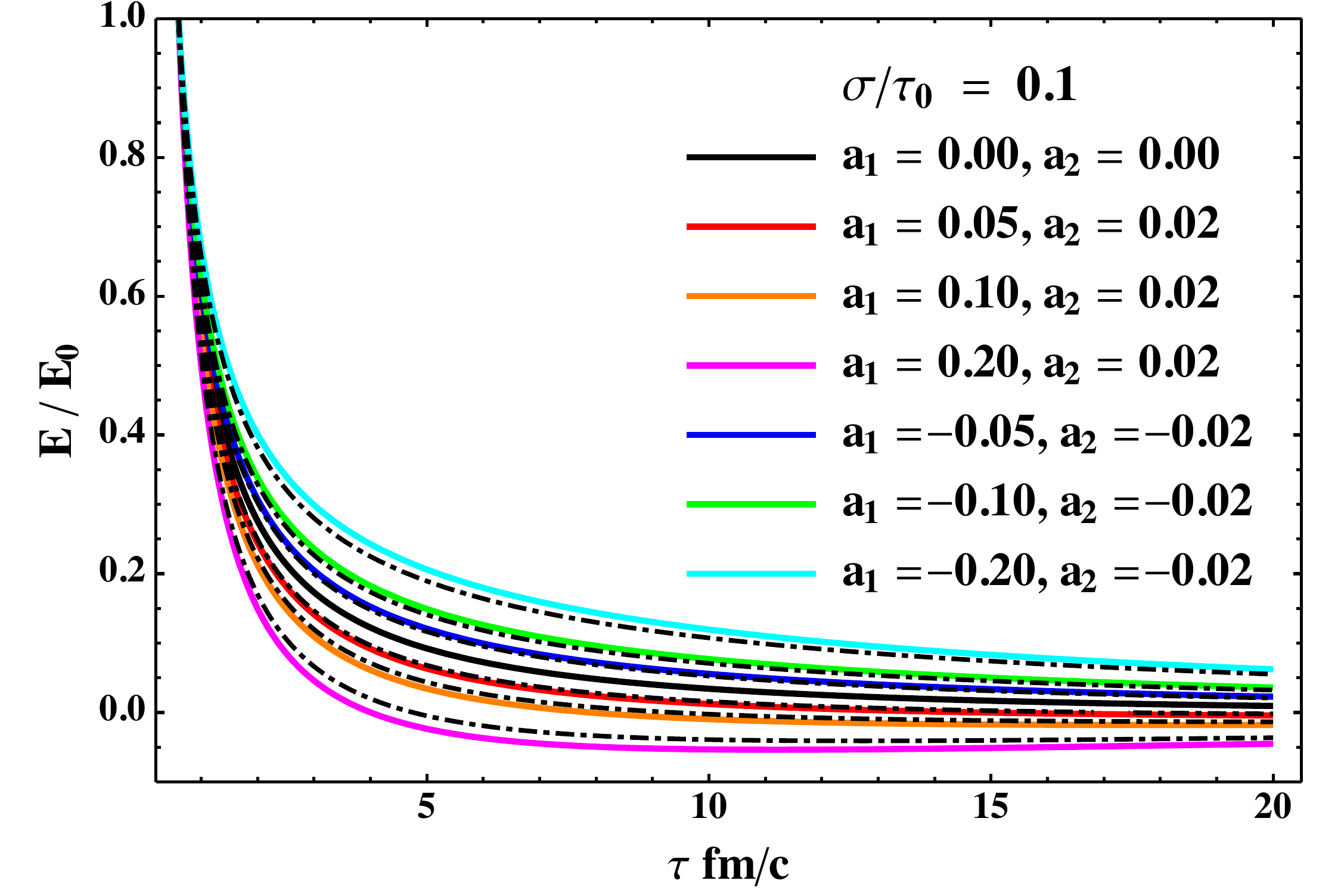}\includegraphics[scale=0.35]{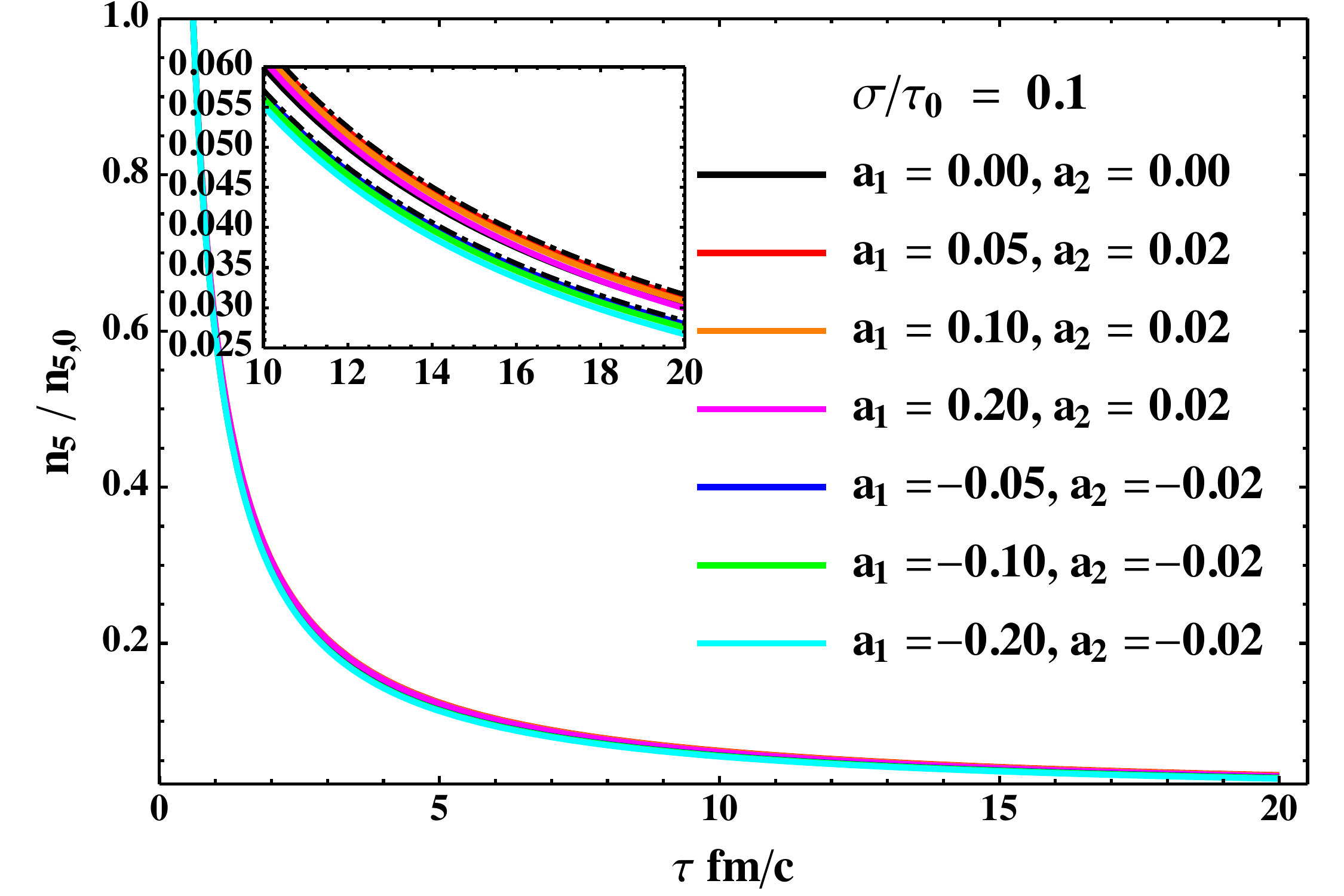}

\includegraphics[scale=0.35]{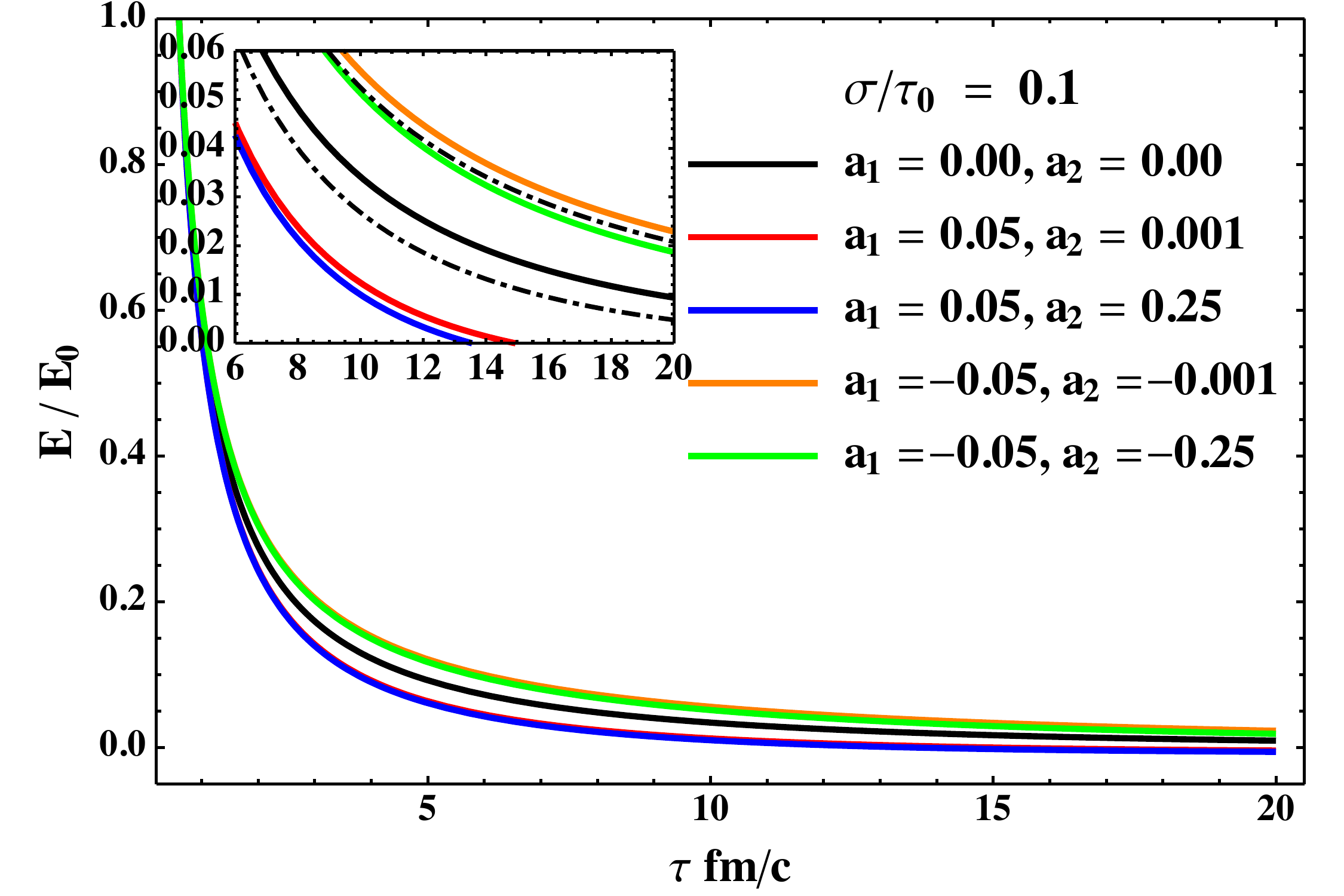}\includegraphics[scale=0.35]{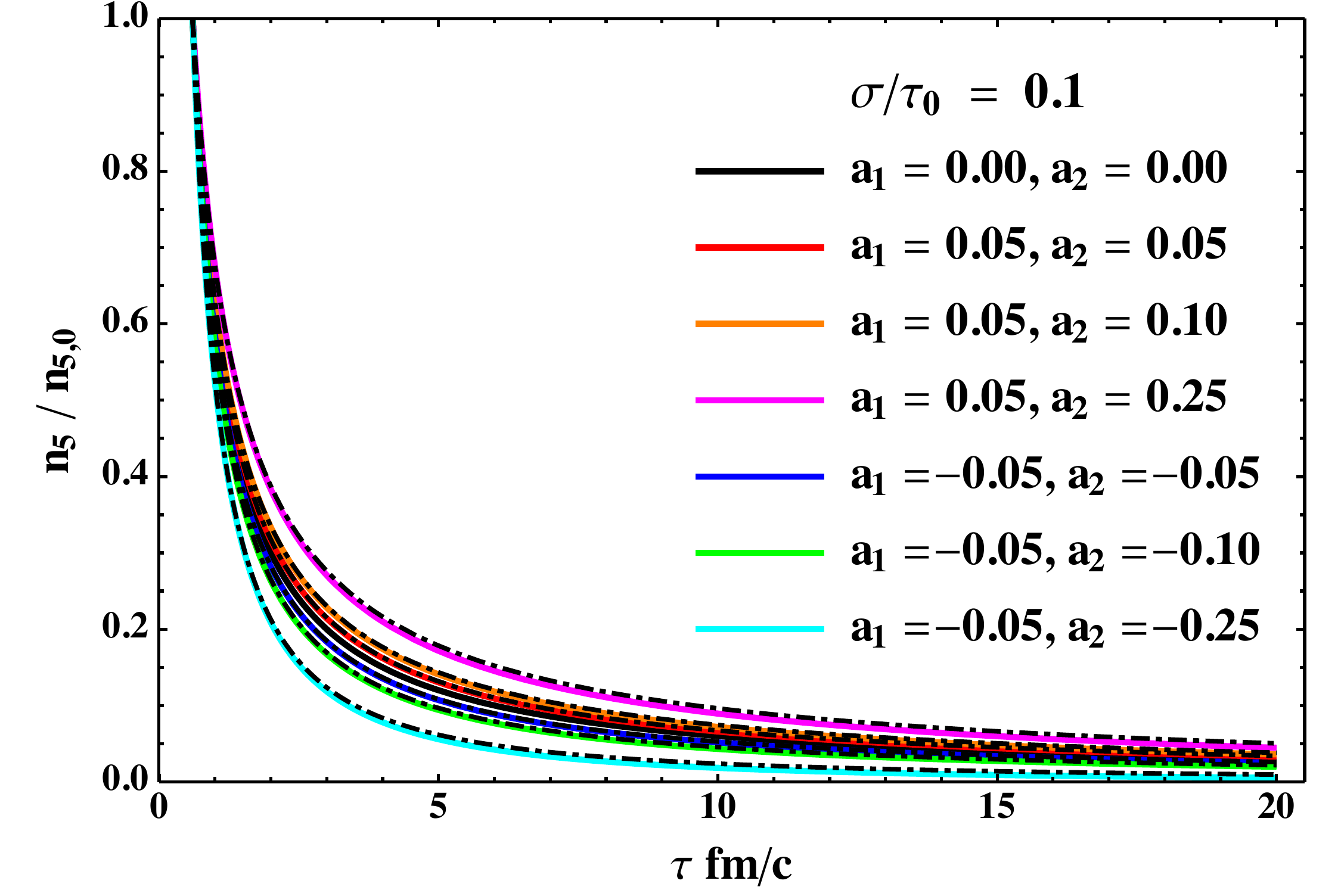}

\includegraphics[scale=0.28]{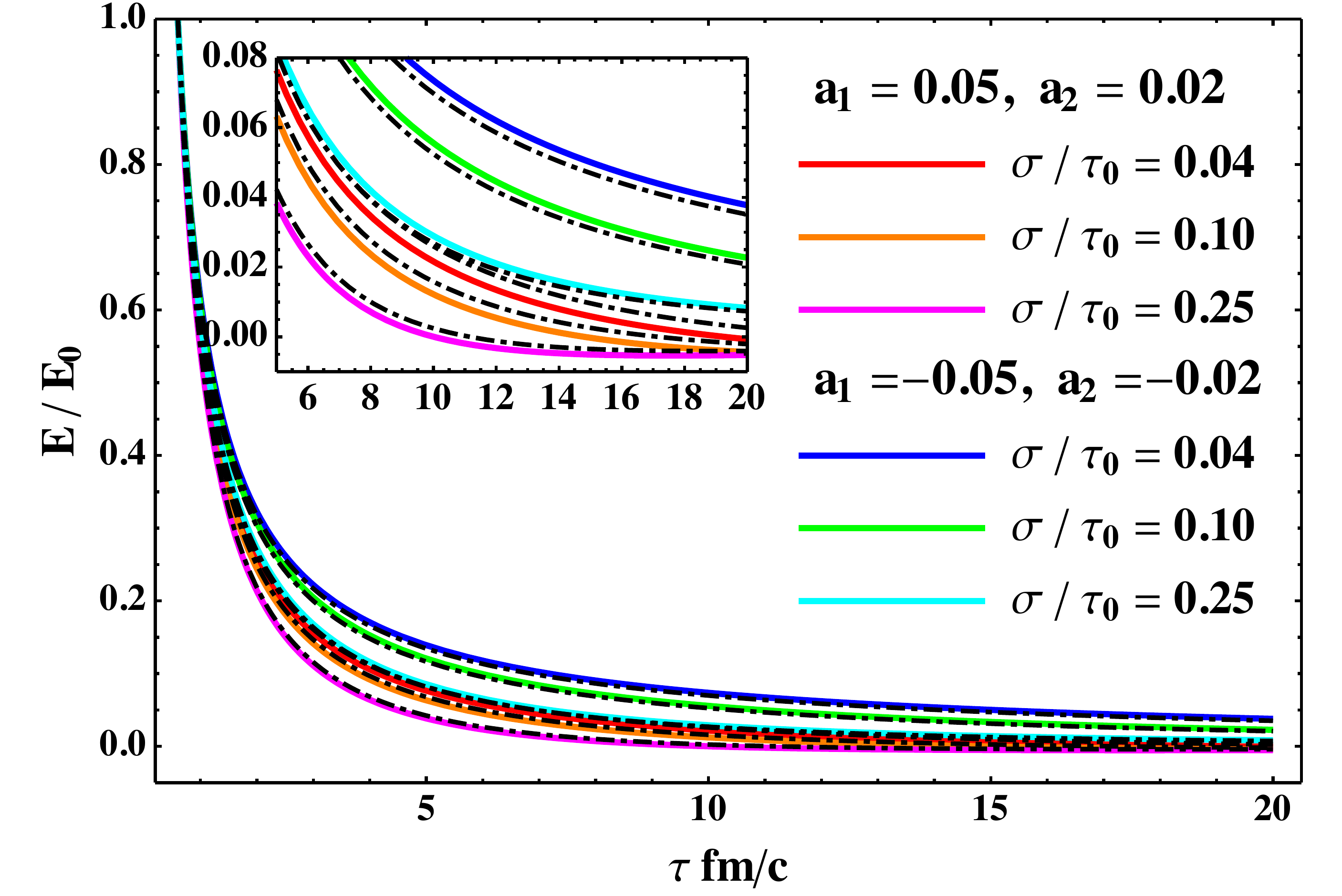}\includegraphics[scale=0.28]{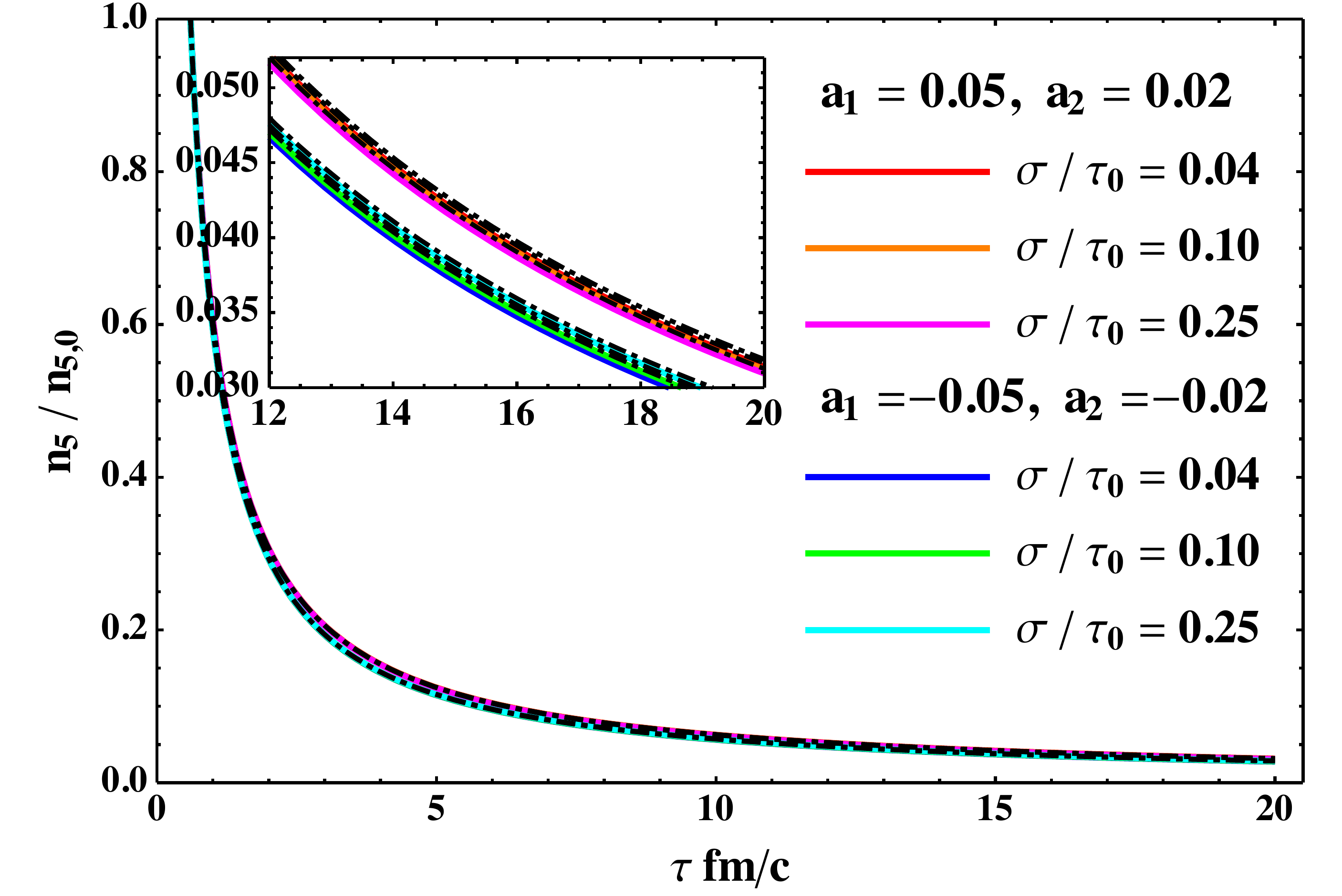}

\caption{The normalized electric field $E/E_{0}$ and chiral charge density
$n_{5}/n_{5,0}$ as functions of the proper time $\tau$. We have
chosen $\tau_{0}=0.6\textrm{ fm/c}$. The solid lines are obtained
by solving Eq. (\ref{eq:xy_01}) numerically and the dashed lines
are from the approximate analytic solution (\ref{eq:EN_A_01}). In
the first row, we fix $\sigma/\tau_{0}=0.1$, $a_{2}=\pm0.2$ and
change the values of $a_{1}$. In the second row, we fix $\sigma/\tau_{0}=0.1$,
$a_{1}=\pm0.5$ and change the values of $a_{2}$. In the last row,
we fix $(a_{1},a_{2})=\pm(0.05,0.02)$ and change the values of $\sigma/\tau_{0}$.
\label{fig:EN_n}}

\end{figure}

We can also solve Eq. (\ref{eq:xy_01}) numerically. We choose the
initial proper time $\tau_{0}=0.6\textrm{ fm/c}$. The values of the
electric conductivity vary in different situations. The lattice QCD
calculations give $\sigma\sim5.8T/T_{c}\textrm{ MeV}$ \citep{Aarts:2007wj,Ding:2010ga,Tuchin:2013ie},
while in holographic QCD models it takes the value $\sigma\sim20-30\textrm{ MeV}$
for $T=200\textrm{ MeV}$ \citep{Pu:2014cwa,Pu:2014fva}. For $\sigma$
in the weakly coupled QGP at finite temperature and chemical potential,
see, e.g. Ref. \citep{Chen:2013tra}. In our numerical calculation,
we choose $\sigma\sim5-30\textrm{ MeV}\simeq0.04-0.25\tau_{0}$. 

\begin{figure}[tp]
\includegraphics[scale=0.28]{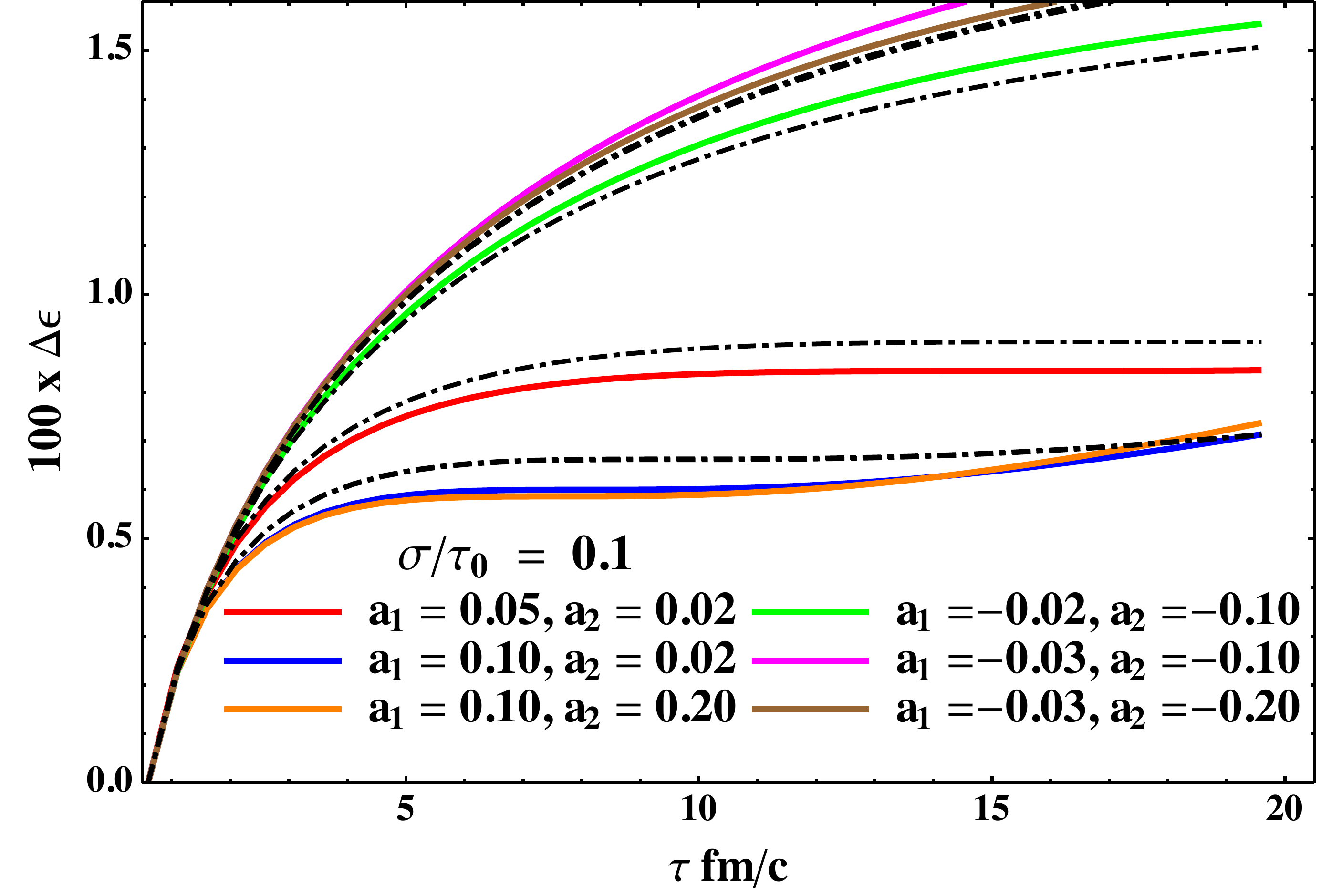}\includegraphics[scale=0.29]{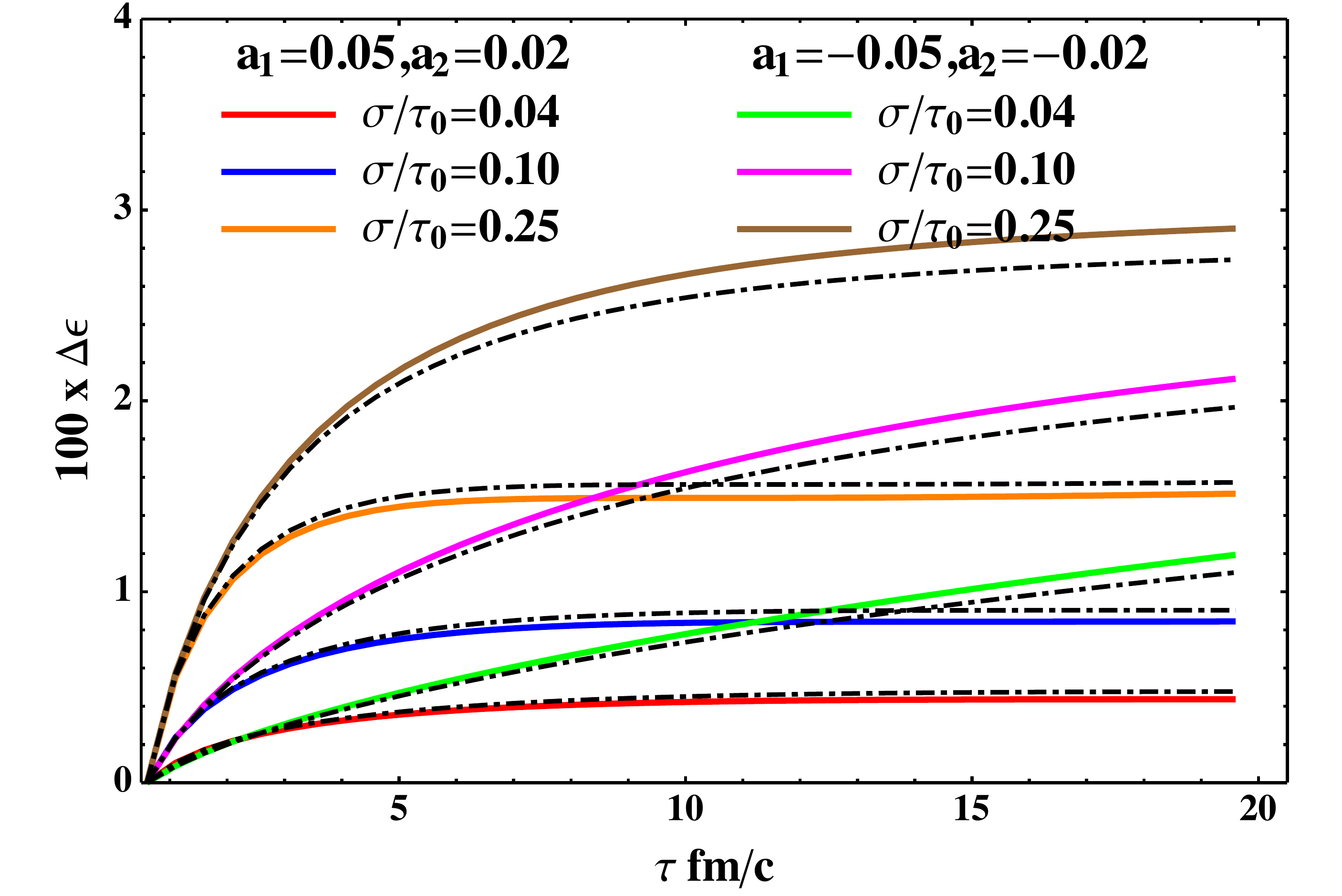}

\caption{The energy density correction $\Delta\varepsilon$ ($\times100$)
as functions the proper time $\tau$. The parameters are set to $\tau_{0}=0.6\textrm{ fm/c}$,
$E_{0}/\epsilon_{0}=0.1$, $B_{0}/\epsilon_{0}=0.2$, $\mu_{5,0}/\tau_{0}=1$,
and $c_{s}^{2}=1/3$. The solid lines are numerical solutions of Eq.
(\ref{eq:xy_01}) and the dashed lines are from the approximate analytic
solution (\ref{eq:EN_A_01}). In the left panel, we fix $\sigma/\tau_{0}=0.1$
and change the values of $a_{1}$ and $a_{2}$. In the right panel,
we fix $(a_{1},a_{2})=\pm(0.05,0.02)$ and change the values of $\sigma/\tau_{0}$.
\label{fig:end_n}}
\end{figure}

In Fig. \ref{fig:EN_n}, we plot the normalized electric field $E/E_{0}$
and chiral charge density $n_{5}/n_{5,0}$ as functions of the proper
time $\tau$. The solid lines are the numerical results from Eqs.
(\ref{eq:xy_01}), while the dashed lines are from the approximate
analytic solution (\ref{eq:EN_A_01}). Note that the approximate analytic
solution for $E(\tau)$ is independent of $a_{2}$ and $n_{5}(\tau)$
independent of $a_{1}$ and $\sigma$. From these results, we see
that the approximation works very well for small $a_{1}$ and $a_{2}$.
For positive $a_{1}$ and $a_{2}$, $E$ decay faster as $a_{1}$
increases, while for negative $a_{1}$ and $a_{2}$, $E$ decay slower
as $|a_{1}|$ increases. For positive $a_{1}$ and $a_{2}$, $n_{5}$
decays slower as $a_{2}$ grows, while for negative $a_{1}$ and $a_{2}$,
$n_{5}$ decays faster as $|a_{2}|$ grows. Such behaviors are obvious
in the approximate analytic solution (\ref{eq:EN_A_01}). 

We observe that for large positive $a_{1}$ or large $\sigma$ with
positive $a_{1}$ and $a_{2}$, $E/E_{0}$ can be negative at late
proper time. It means that the electric field flips its sign at the
late time. From Eq. (\ref{eq:EN_A_01}), one can see that a very large
$a_{1}$ in the second term may dominate and make $E/E_{0}$ negative.
Since $a_{1}$ is proportional to the initial chiral charge density,
such a behavior may come from the competition between the anomalous
conservation equation $\partial_{\mu}j_{5}^{\mu}=-CE\cdot B$ and
Maxwell's equations.

One may expect that $n_{5}$ may have oscillation with time because
it can be converted from the magnetic helicity and vice versa \citep{Akamatsu2013}.
However, since the medium is expanding, the possible oscillation of
$n_{5}$ is outperformed by its decay $n_{5}/n_{5,0}\sim\tau_{0}/\tau$. 

In Fig. \ref{fig:end_n}, we show the results of $\Delta\varepsilon$
in Eq. (\ref{eq:end_01}) which is amplified by a factor 100. The
solid lines are numerical results from Eq. (\ref{eq:xy_01}), while
the dashed lines are given by the approximate analytic solution (\ref{eq:EN_A_01}).
Even with 100 times amplification of the difference, we see that the
approximate analytic solution (\ref{eq:EN_A_01}) still works well.
For both positive and negative $a_{1}$ and $a_{2}$, $\Delta\varepsilon$
are positive because the first term dominates over the second one
inside the square brackets in Eq. (\ref{eq:end_01}).

\subsection{EoS (\ref{eq:eos_02})}

\label{subsec:For-EoS-2}For EoS (\ref{eq:eos_02}), the equations
for the energy density $\varepsilon(\tau)$, $E(\tau)$ and $n_{5}(\tau)$
are coupled together. We need to rewrite Eqs. (\ref{eq:energy_density_03},
\ref{eq:Maxwell_02a}, \ref{eq:conserved_current_02}) as 
\begin{eqnarray}
\frac{d}{d\tau}\varepsilon+(1+c_{s}^{2})\varepsilon & = & \varepsilon\frac{d}{d\tau}\mathcal{L},\nonumber \\
\frac{d}{d\tau}E+\frac{E}{\tau} & = & E\frac{d}{d\tau}\mathcal{E},\nonumber \\
\frac{d}{d\tau}n_{5}+\frac{n_{5}}{\tau} & = & n_{5}\frac{d}{d\tau}\mathcal{N},\label{eq:EN_04-1}
\end{eqnarray}
where 
\begin{eqnarray}
\frac{d}{d\tau}\mathcal{L} & = & \frac{1}{\varepsilon}\sigma E^{2}+\frac{1}{\varepsilon}eC\chi\mu_{5}EB,\nonumber \\
\frac{d}{d\tau}\mathcal{E} & = & -\sigma-eC\chi\mu_{5}\frac{B}{E},\nonumber \\
\frac{d}{d\tau}\mathcal{N} & = & \frac{e^{2}C\chi EB}{n_{5}}.\label{eq:EN_03-1}
\end{eqnarray}
With the help of Eq. (\ref{eq:general_form_01}), the solutions are,
\begin{equation}
\varepsilon(\tau)=\varepsilon_{0}\left(\frac{\tau_{0}}{\tau}\right)^{1+c_{s}^{2}}\exp\left[\mathcal{L}(\tau)-\mathcal{L}(\tau_{0})\right],
\end{equation}
and $n_{5}(\tau)$ and $E(\tau)$ are similar to Eq. (\ref{eq:nE_00}).

From the EoS (\ref{eq:eos_02}), one can express all thermodynamic
quantities as functions of $T$ and $\mu_{5}$. Since the critical
temperature $T_{c}\sim200\textrm{ MeV}$ is much larger than the chiral
chemical potential in relativistic heavy ion collisions, i.e. $\mu_{5}\ll T$,
all terms proportional to $\mu_{5}$ in the thermodynamic relations
are negligible. As a consequence, we obtain 
\begin{equation}
\varepsilon=\varepsilon_{0}\left(\frac{T}{T_{0}}\right)^{1+c_{s}^{-2}}+\mathcal{O}(\mu_{5}^{2}/T^{2}),
\end{equation}
where $\varepsilon_{0}=\varepsilon(\tau_{0})$ and $T_{0}=T(\tau_{0})$.
By introducing,
\begin{eqnarray}
x(\tau) & = & \exp\left[\mathcal{E}(\tau)-\mathcal{E}(\tau_{0})\right],\nonumber \\
y(\tau) & = & \exp\left[\mathcal{N}(\tau)-\mathcal{N}(\tau_{0})\right],\nonumber \\
z(\tau) & = & \exp\left[\mathcal{L}(\tau)-\mathcal{L}(\tau_{0})\right],
\end{eqnarray}
Equation (\ref{eq:EN_03-1}) is reduced to 
\begin{eqnarray}
\frac{d}{d\tau}x & = & -\sigma x-\frac{a_{1}}{\tau_{0}}y(\tau)\left(\frac{\tau}{\tau_{0}}\right)^{-1+2c_{s}^{2}}z^{-2c_{s}^{2}/(1+c_{s}^{2})},\nonumber \\
\frac{d}{d\tau}y & = & a_{2}\frac{x(\tau)}{\tau},\nonumber \\
\frac{d}{d\tau}z & = & \sigma\frac{E_{0}^{2}}{\varepsilon_{0}}\left(\frac{\tau_{0}}{\tau}\right)^{1-c_{s}^{2}}x^{2}(\tau)+\frac{a_{3}}{\tau_{0}}\left(\frac{\tau}{\tau_{0}}\right)^{-2+3c_{s}^{2}}x(\tau)y(\tau)z^{-2c_{s}^{2}/(1+c_{s}^{2})},\label{eq:xyz_01}
\end{eqnarray}
where $x(\tau_{0})=y(\tau_{0})=z(\tau_{0})=1$, and $a_{1}$, $a_{2}$
and $a_{3}$ are dimensionless constants determined by the initial
conditions 
\begin{eqnarray}
a_{1} & = & eC\chi\frac{B_{0}n_{5,0}}{aT_{0}^{2}E_{0}}\tau_{0},\nonumber \\
a_{2} & = & \frac{e^{2}C\chi E_{0}B_{0}}{n_{5,0}}\tau_{0},\nonumber \\
a_{3} & = & \frac{eC\chi}{a}\frac{n_{5,0}E_{0}B_{0}}{\varepsilon_{0}T_{0}^{2}}\tau_{0}.
\end{eqnarray}
These dimensionless constants are all linearly proportional to $\hbar$
through the anomaly constant $C$, which means they are of quantum
nature. So we can deal with the terms proportional to $a_{1},$ $a_{2}$
and $a_{3}$ in Eq.(\ref{eq:xyz_01}) as perturbations to the classical
terms, and Eq. (\ref{eq:xyz_01}) can be solved order by order in
powers of $\hbar$. 

\begin{figure}
\includegraphics[scale=0.28]{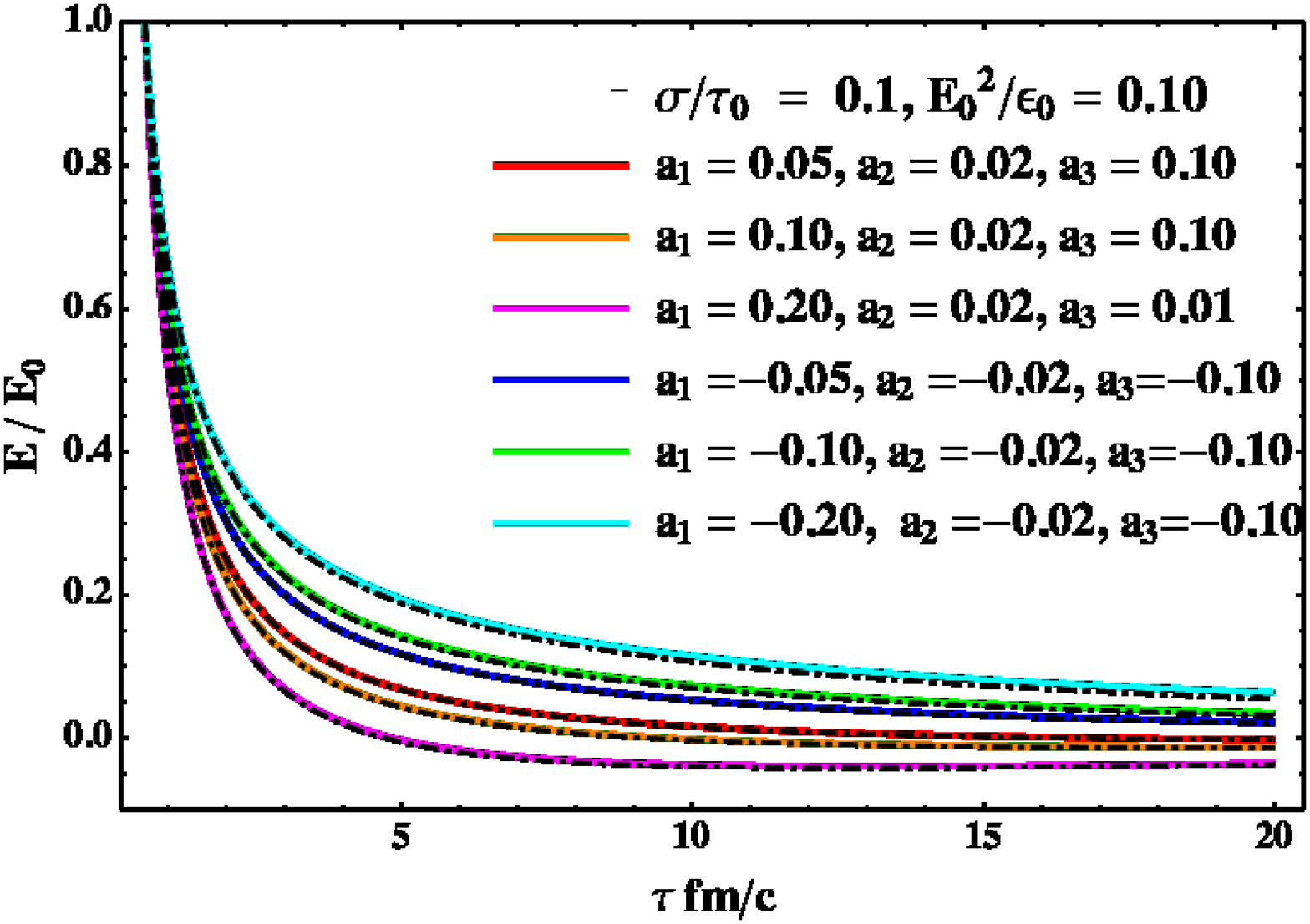}\includegraphics[scale=0.28]{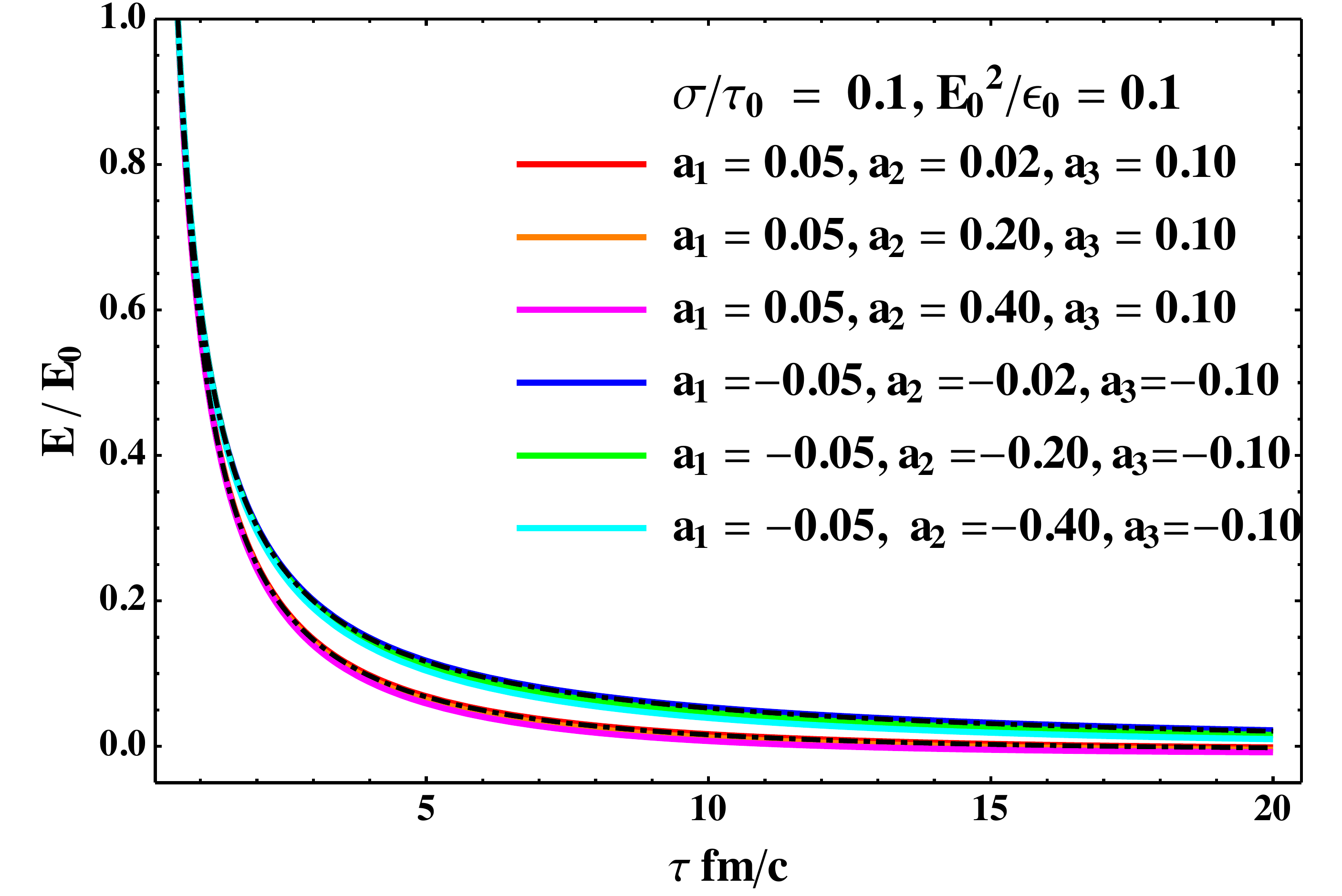}

\includegraphics[scale=0.28]{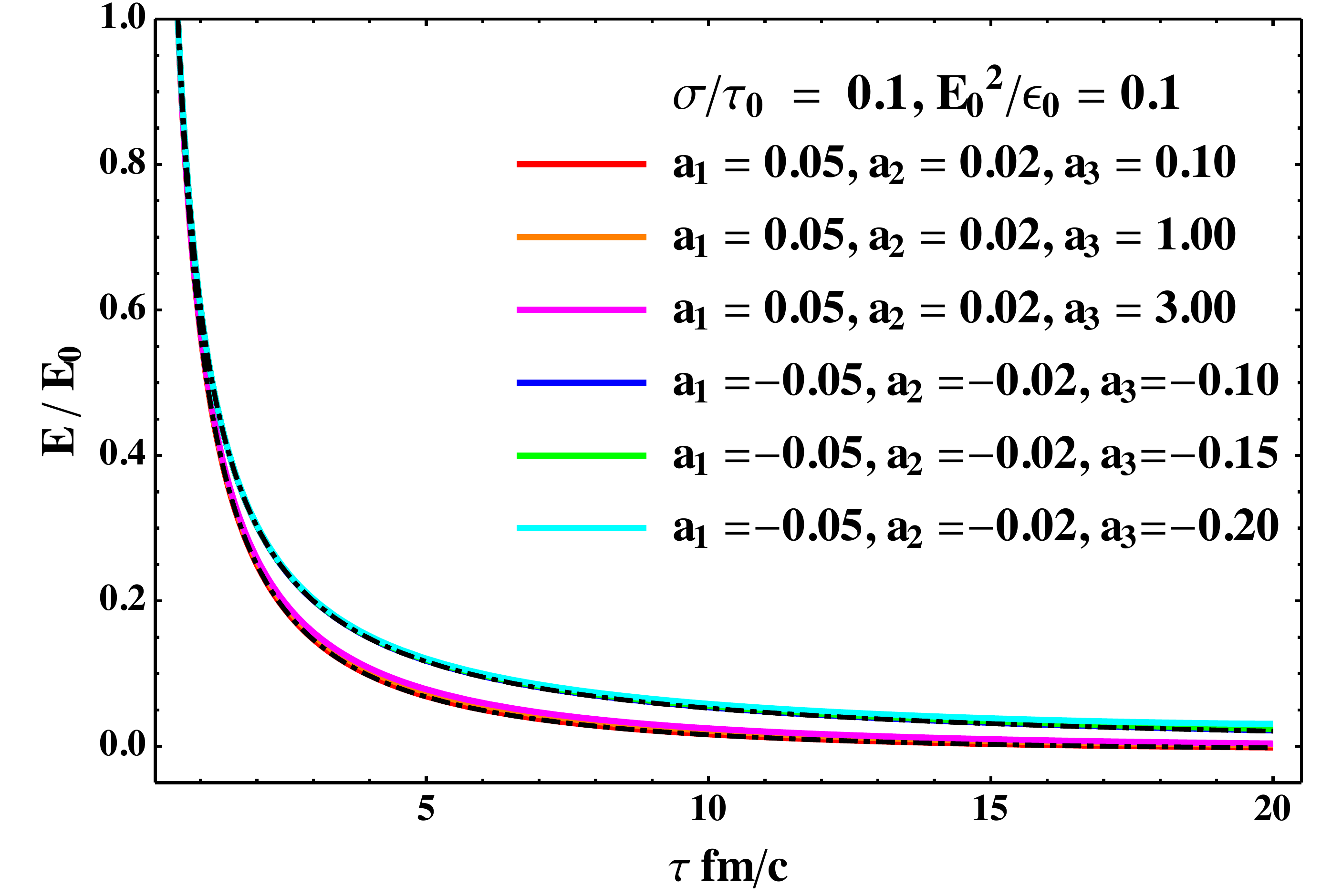}\includegraphics[scale=0.28]{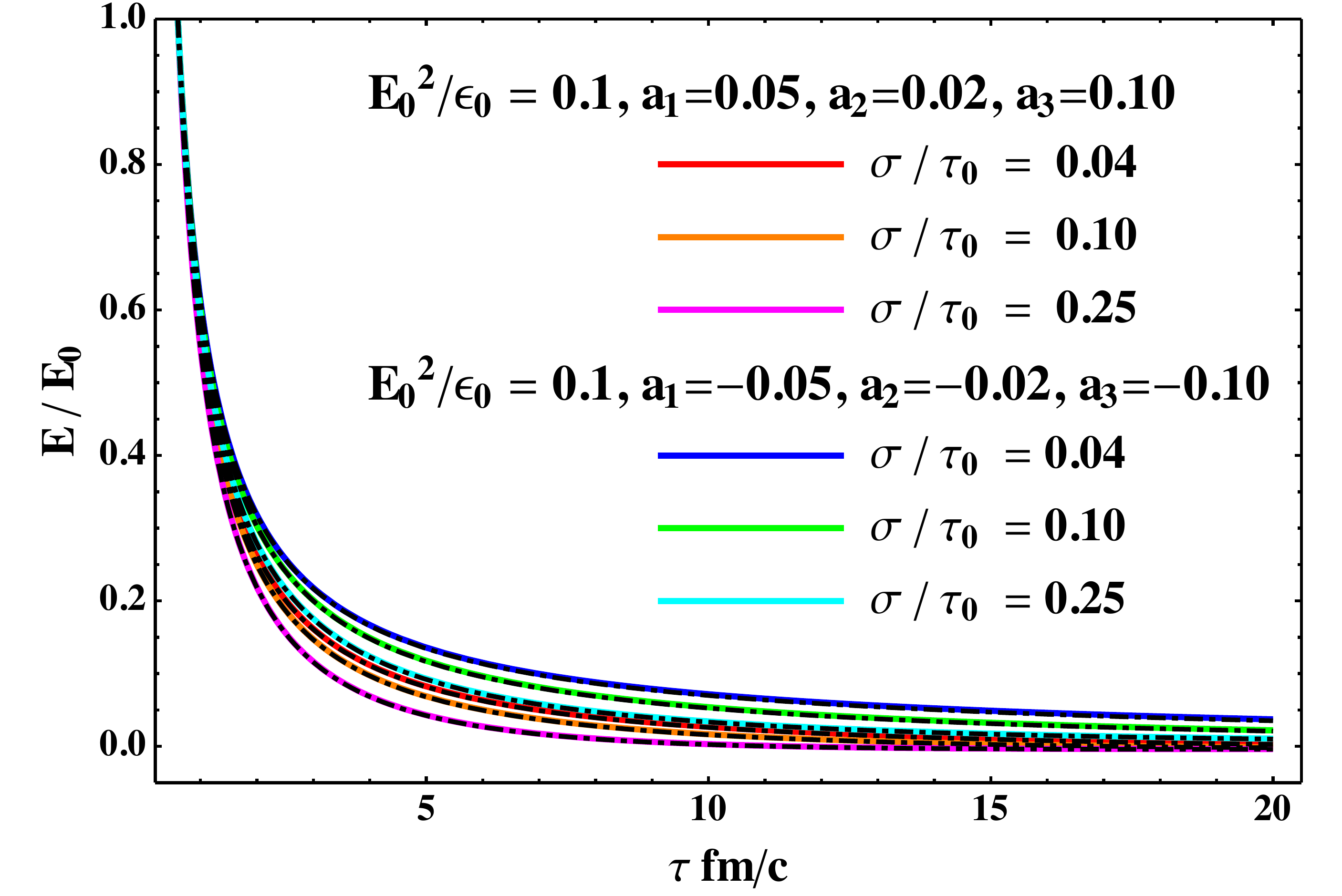}

\caption{The normalized electric field $E/E_{0}$ as functions of the proper
time $\tau$. We have chosen $\tau_{0}=0.6\textrm{ fm/c}$, $c_{s}^{2}=1/3$
and $E_{0}^{2}/\varepsilon_{0}=0.1$. The solid lines are obtained
by solving Eq. (\ref{eq:xyz_01}) numerically and the dashed lines
are from the approximate analytic solution (\ref{eq:sol_02}). In
the first row, we fix $\sigma/\tau_{0}=0.1,$ $a_{2}=\pm0.2$, $a_{3}=\pm0.10$
and change the values of $a_{1}$. In the second row, we fix $\sigma/\tau_{0}=0.1$,
$a_{1}=\pm0.5$ and change the values of $a_{2}$. In the last row,
we fix $(a_{1},a_{2})=\pm(0.05,0.02)$ and change the values of $\sigma/\tau_{0}$.
\label{fig:ET}}
\end{figure}

To the linear order in $\hbar$, we have the solutions for $x(\tau)$,
$y(\tau)$ and $z(\tau)$

\begin{eqnarray}
x(\tau) & = & e^{-\sigma(\tau-\tau_{0})}-\frac{a_{1}}{\tau_{0}}e^{-\sigma\tau}\int_{\tau_{0}}^{\tau}d\tau^{\prime}e^{\sigma\tau^{\prime}}\left(\frac{\tau^{\prime}}{\tau_{0}}\right)^{-1+2c_{s}^{2}}[z_{0}(\tau^{\prime})]^{-2c_{s}^{2}/(1+c_{s}^{2})},\nonumber \\
y(\tau) & = & 1+a_{2}e^{\sigma\tau_{0}}[\textrm{E}_{1}(\sigma\tau_{0})-\textrm{E}_{1}(\sigma\tau)],\nonumber \\
z(\tau) & = & z_{0}(\tau)+\frac{a_{3}}{\tau_{0}}\int_{\tau_{0}}^{\tau}d\tau^{\prime}\left(\frac{\tau^{\prime}}{\tau_{0}}\right)^{-2+3c_{s}^{2}}e^{-\sigma(\tau^{\prime}-\tau_{0})}[z_{0}(\tau^{\prime})]^{-2c_{s}^{2}/(1+c_{s}^{2})},
\end{eqnarray}
where 
\begin{equation}
z_{0}(\tau)=1+\sigma\frac{E_{0}^{2}}{\varepsilon_{0}}e^{2\sigma\tau_{0}}[\tau_{0}\textrm{E}_{1-c_{s}^{2}}(2\sigma\tau_{0})-\tau\left(\frac{\tau}{\tau_{0}}\right)^{c_{s}^{2}-1}\textrm{E}_{1-c_{s}^{2}}(2\sigma\tau^{\prime})].
\end{equation}
We can further simplify the integration in $x(\tau)$ and $z(\tau)$.
Since initial energy density $\varepsilon_{0}$ is much larger than
the initial energy of the EM fields $\varepsilon_{0}\gg B_{0}^{2},E_{0}^{2},E_{0}B_{0}$
(see, e.g., Ref. \citep{Roy:2015coa} for the values of $B_{0}^{2}/\varepsilon_{0}$
in the event-by-event simulation of relativistic heavy ion collisions),
we can further simplify the integration in $x(\tau)$ and $z(\tau)$
in the linear order in $E_{0}^{2}/\varepsilon_{0}$ as
\begin{eqnarray}
x(\tau) & = & e^{-\sigma(\tau-\tau_{0})}-\frac{a_{1}}{\tau_{0}}e^{-\sigma\tau}[\tau_{0}\textrm{E}_{1-2c_{s}^{2}}(-\sigma\tau_{0})-\tau\left(\frac{\tau}{\tau_{0}}\right)^{-1+2c_{s}^{2}}\textrm{E}_{1-2c_{s}^{2}}(-\sigma\tau)]+\mathcal{O}(a_{i}^{2},a_{i}E_{0}^{2}/\varepsilon_{0}),\nonumber \\
z(\tau) & = & 1+\sigma\frac{E_{0}^{2}}{\varepsilon_{0}}e^{2\sigma\tau_{0}}[\tau_{0}\textrm{E}_{1-c_{s}^{2}}(2\sigma\tau_{0})-\tau\left(\frac{\tau}{\tau_{0}}\right)^{c_{s}^{2}-1}\textrm{E}_{1-c_{s}^{2}}(2\sigma\tau)]\nonumber \\
 &  & +\frac{a_{3}}{\tau_{0}}e^{\sigma\tau_{0}}[\tau_{0}\textrm{E}_{2-3c_{s}^{2}}(\sigma\tau_{0})-\tau\left(\frac{\tau_{0}}{\tau}\right)^{2-3c_{s}^{2}}\textrm{E}_{2-3c_{s}^{2}}(\sigma\tau)]+\mathcal{O}(a_{i}^{2},a_{i}E_{0}^{2}/\varepsilon_{0}).
\end{eqnarray}

\begin{figure}
\includegraphics[scale=0.35]{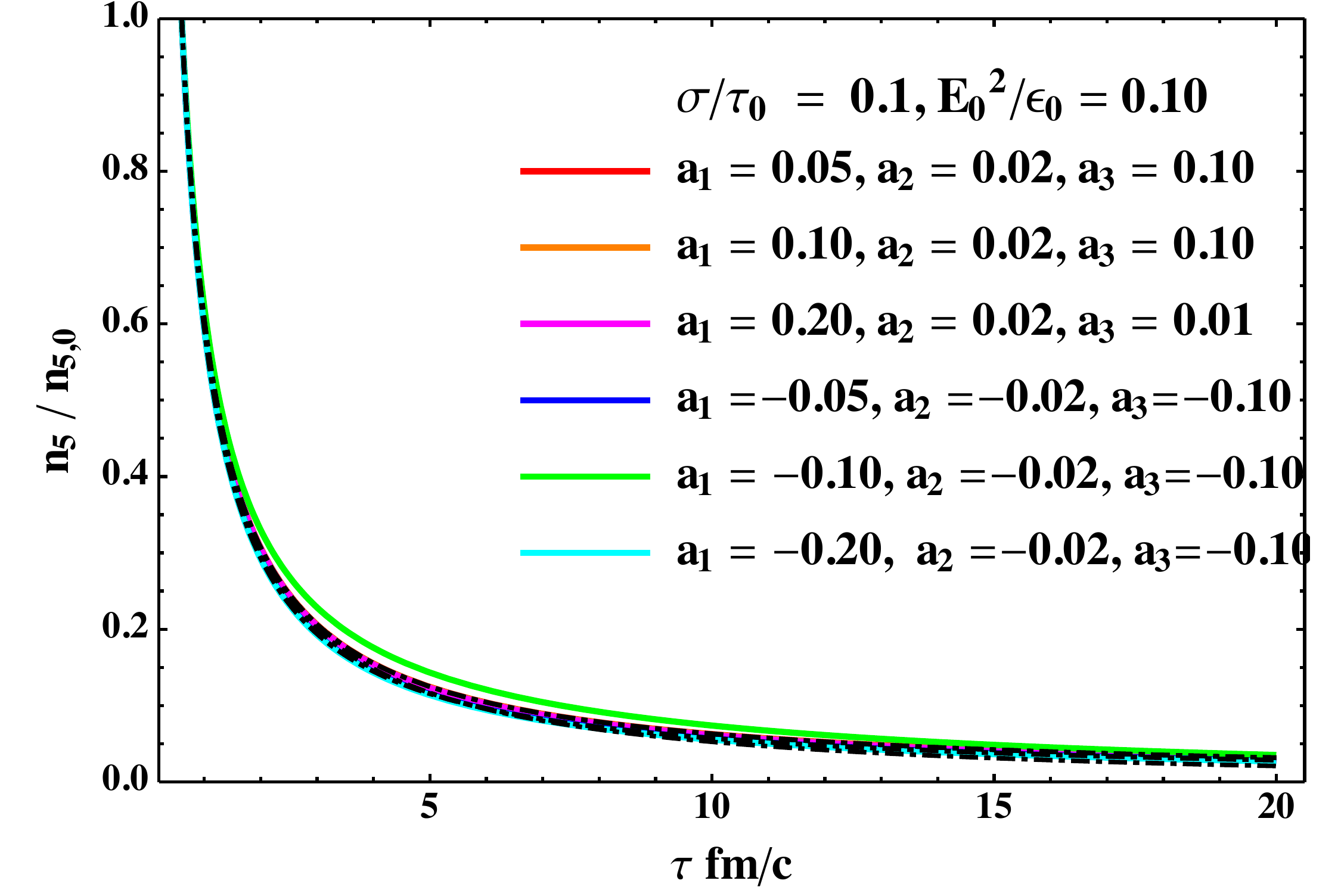}\includegraphics[scale=0.35]{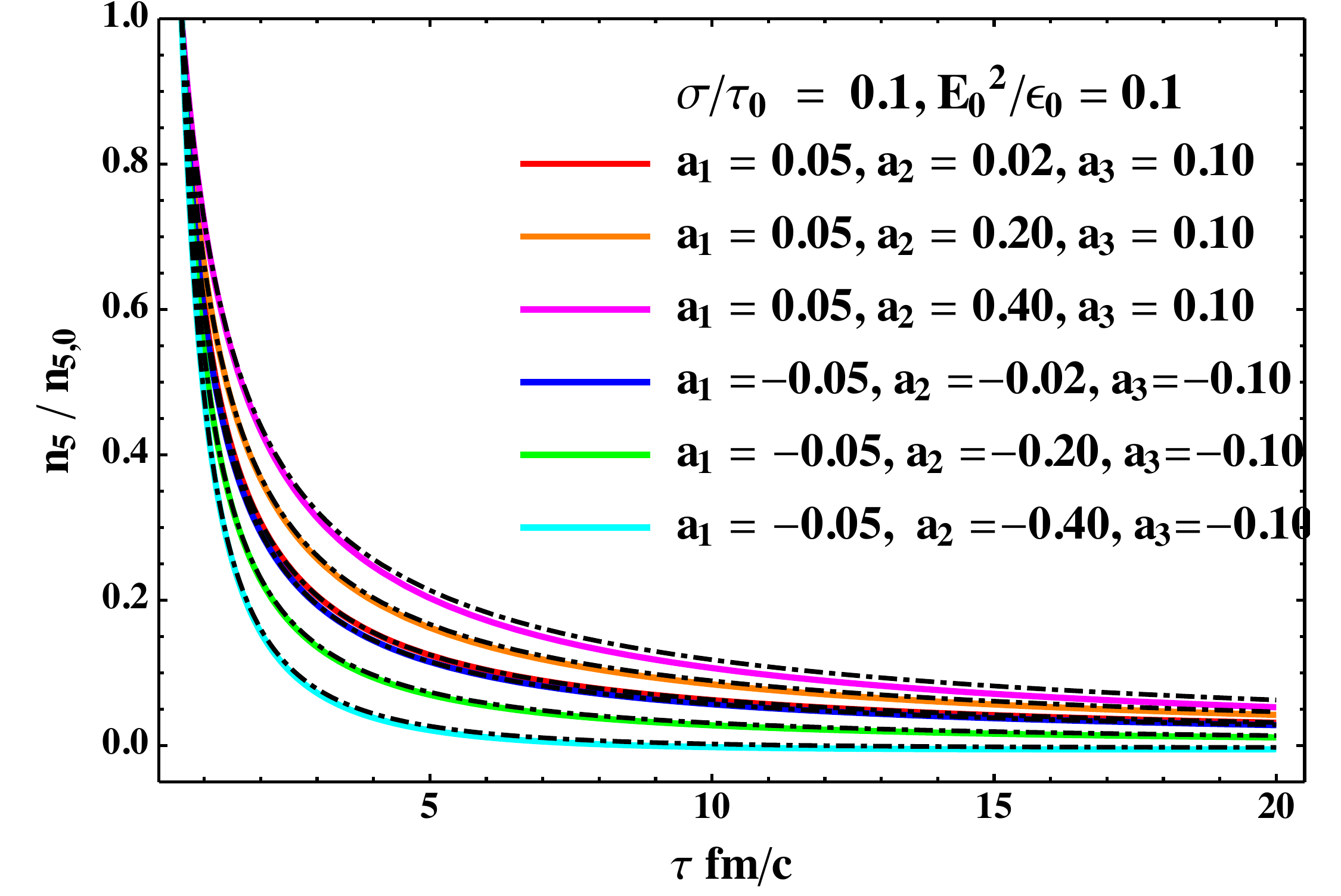}

\includegraphics[scale=0.35]{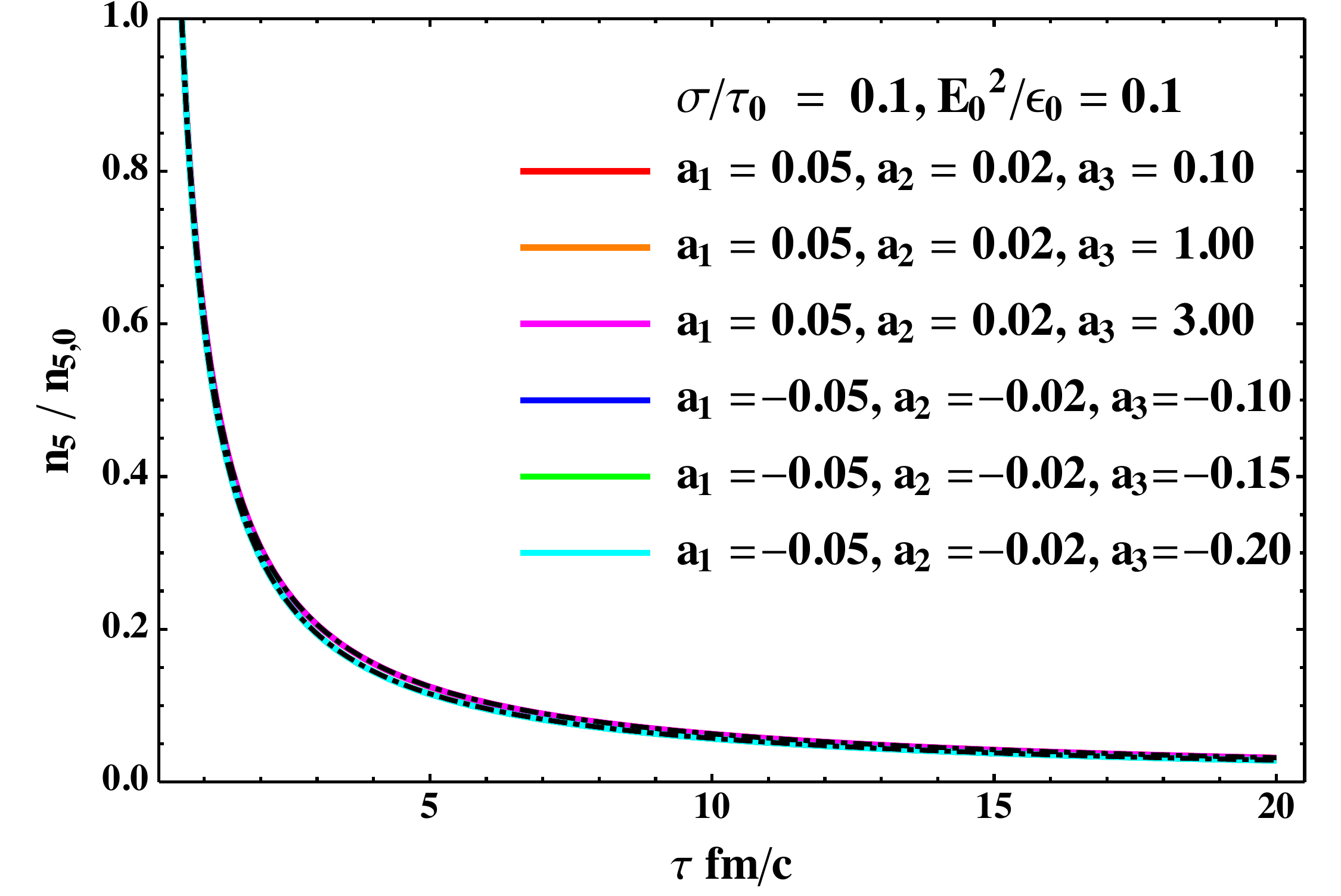}\includegraphics[scale=0.28]{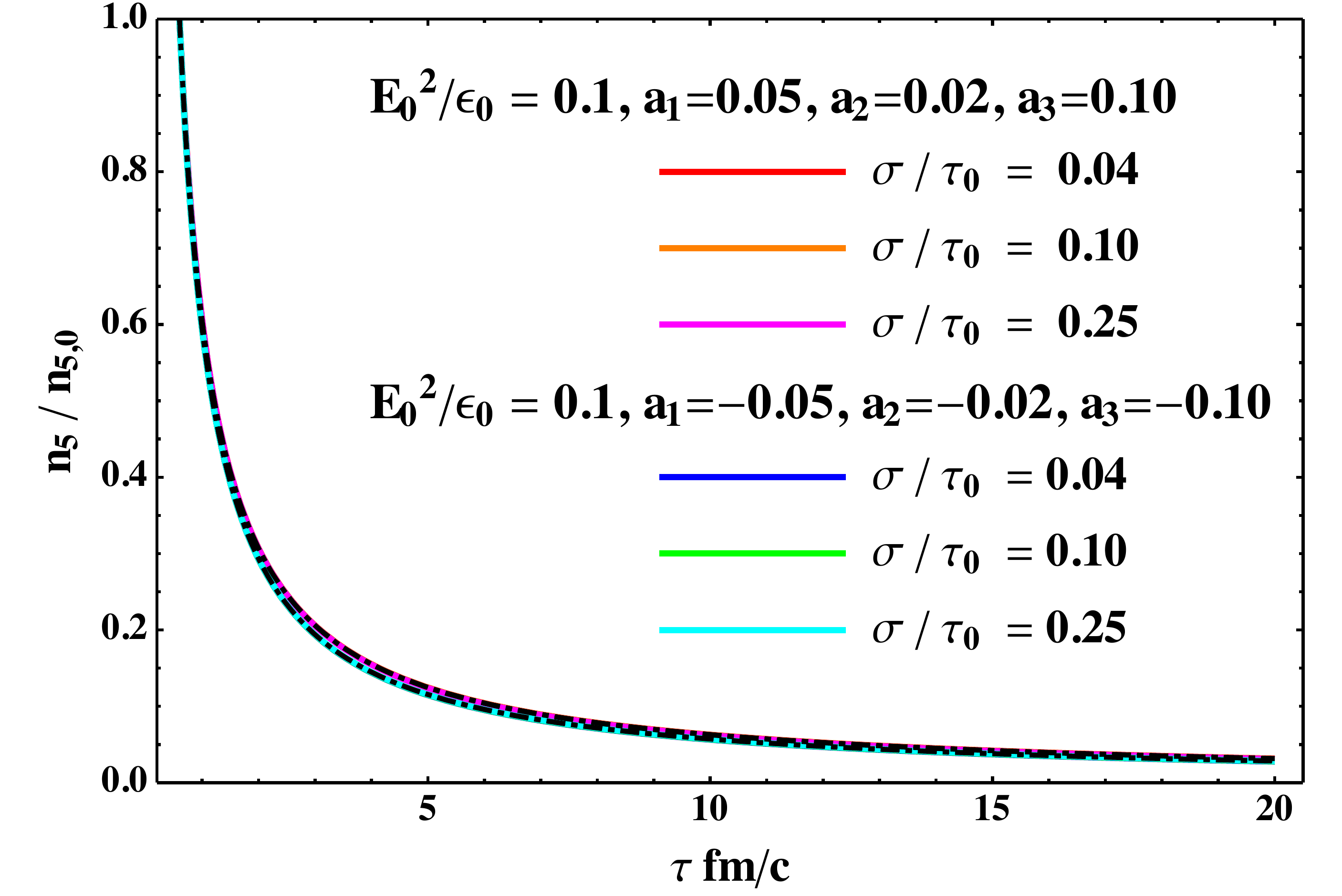}

\caption{The normalized electric field $n_{5}/n_{5,0}$ as functions of the
proper time $\tau$. We have chosen $\tau_{0}=0.6\textrm{ fm/c}$,
$c_{s}^{2}=1/3$ and $E_{0}^{2}/\varepsilon_{0}=0.1$. The solid lines
are obtained by solving Eq. (\ref{eq:xyz_01}) numerically and the
dashed lines are from the approximate analytic solution (\ref{eq:sol_02}).
In the first row, we fix $\sigma/\tau_{0}=0.1,$ $a_{2}=\pm0.2$,
$a_{3}=\pm0.10$ and change the values of $a_{1}$. In the second
row, we fix $\sigma/\tau_{0}=0.1$, $a_{1}=\pm0.5$ and change the
values of $a_{2}$. \label{fig:NT_1}}
\end{figure}

\begin{figure}
\includegraphics[scale=0.28]{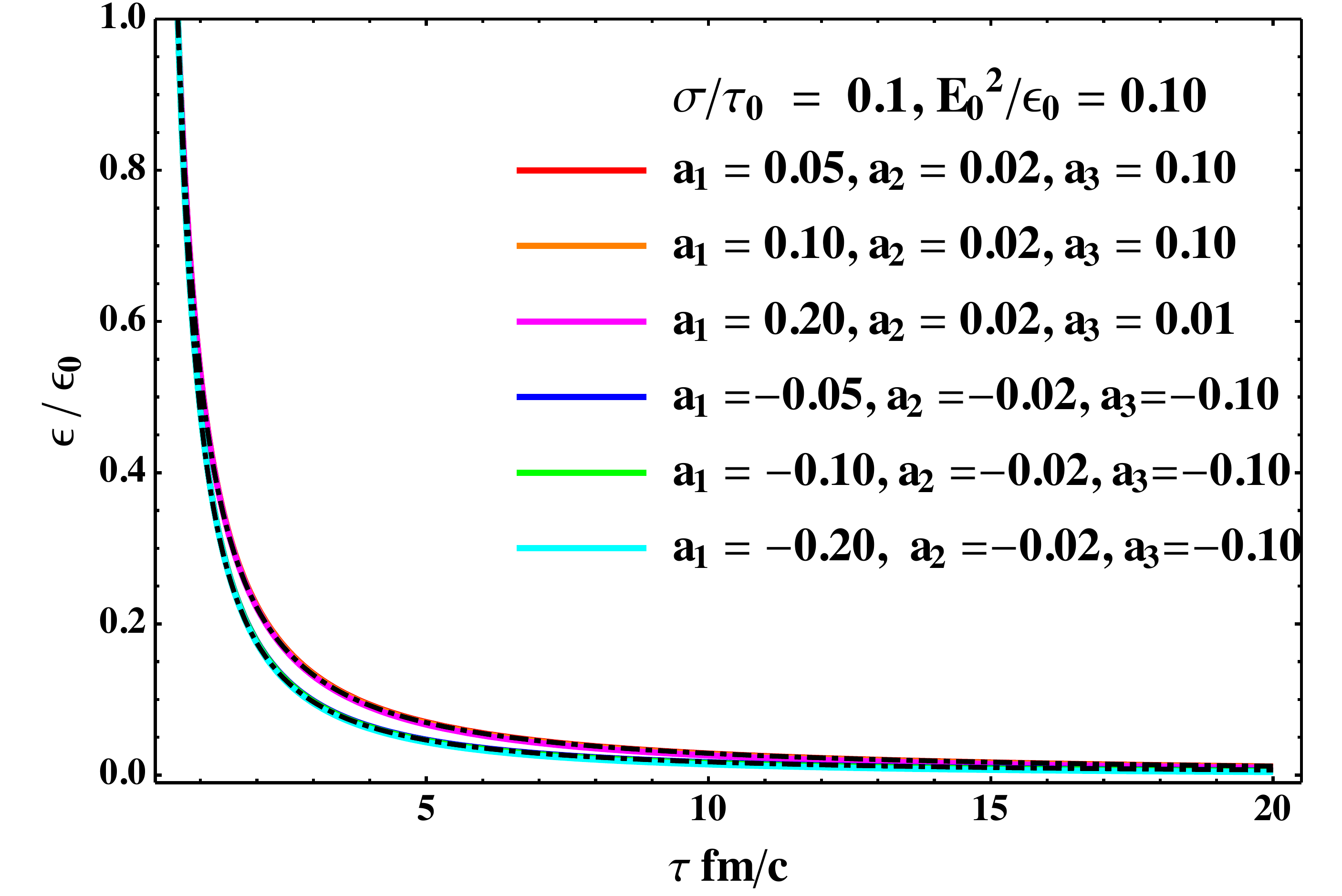}\includegraphics[scale=0.28]{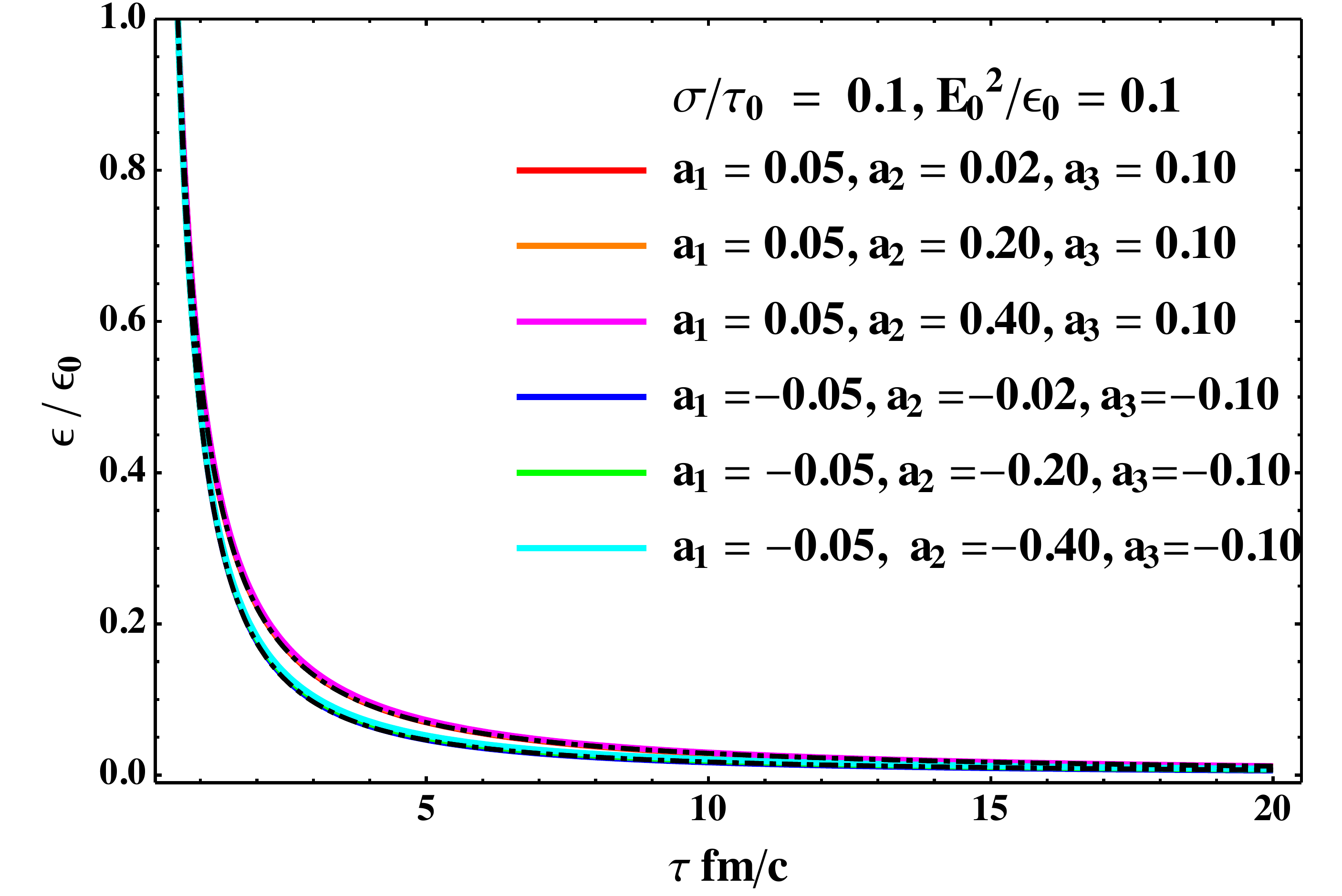}

\includegraphics[scale=0.28]{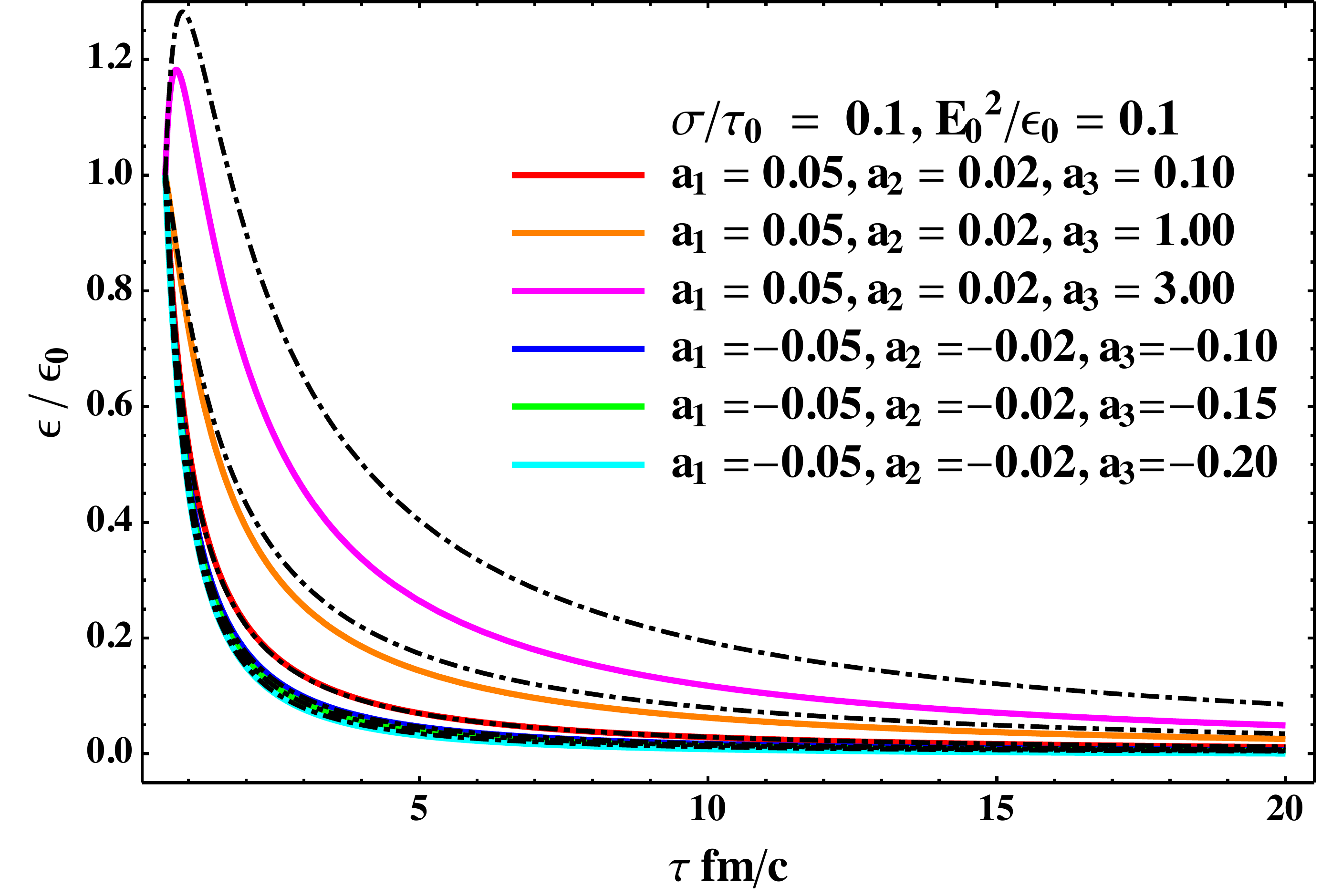} \includegraphics[scale=0.28]{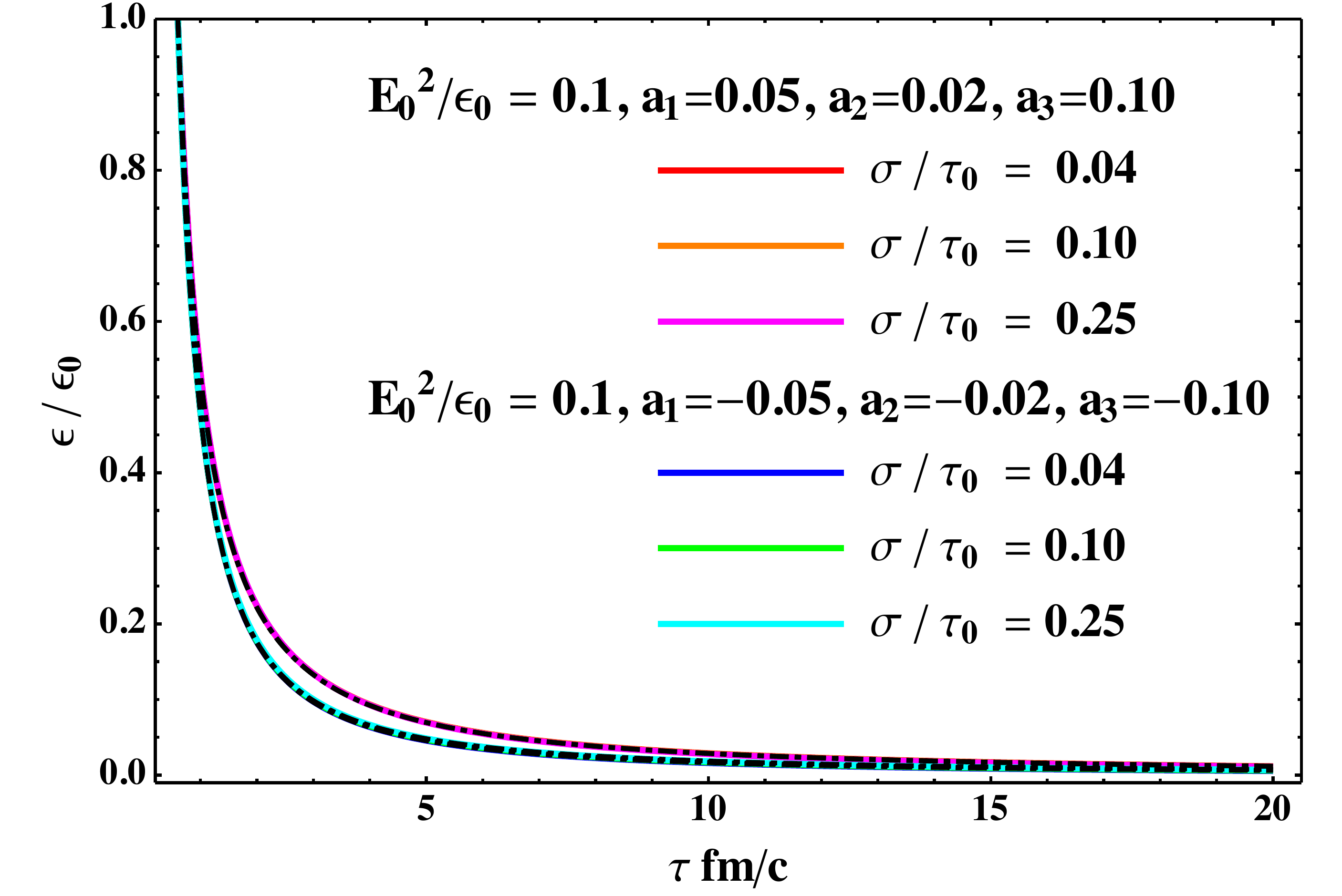}

\caption{The normalized electric field $\varepsilon/\varepsilon_{0}$ as functions
of the proper time $\tau$. We have chosen $\tau_{0}=0.6\textrm{ fm/c}$,
$c_{s}^{2}=1/3$ and $E_{0}^{2}/\varepsilon_{0}=0.1$. The solid lines
are obtained by solving Eq. (\ref{eq:xyz_01}) numerically and the
dashed lines are from the approximate analytic solution (\ref{eq:sol_02}).
In the first row, we fix $\sigma/\tau_{0}=0.1,$ $a_{2}=\pm0.2$,
$a_{3}=\pm0.10$ and change the values of $a_{1}$. In the second
row, we fix $\sigma/\tau_{0}=0.1$, $a_{1}=\pm0.5$ and change the
values of $a_{2}$. \label{fig:endT_1}}
\end{figure}

Then we obtain the solutions for $E(\tau)$ , $n_{5}(\tau)$ and $\varepsilon(\tau)$
in the linear order in $\hbar$ and $E_{0}^{2}/\varepsilon_{0}$
\begin{eqnarray}
E(\tau) & = & E_{0}\left(\frac{\tau_{0}}{\tau}\right)\left\{ e^{-\sigma(\tau-\tau_{0})}-a_{1}e^{-\sigma\tau}[\textrm{E}_{1-2c_{s}^{2}}(-\sigma\tau_{0})-\left(\frac{\tau}{\tau_{0}}\right)^{2c_{s}^{2}}\textrm{E}_{1-2c_{s}^{2}}(-\sigma\tau)]\right\} ,\nonumber \\
n_{5}(\tau) & = & n_{5,0}\left(\frac{\tau_{0}}{\tau}\right)\left\{ 1+a_{2}e^{\sigma\tau_{0}}[\textrm{E}_{1}(\sigma\tau_{0})-\textrm{E}_{1}(\sigma\tau)]\right\} ,\nonumber \\
\varepsilon(\tau) & = & \epsilon_{0}\left(\frac{\tau_{0}}{\tau}\right)^{1+c_{s}^{2}}\left\{ 1+\sigma\frac{E_{0}^{2}}{\varepsilon_{0}}e^{2\sigma\tau_{0}}[\tau_{0}\textrm{E}_{1-c_{s}^{2}}(2\sigma\tau_{0})-\tau\left(\frac{\tau}{\tau_{0}}\right)^{c_{s}^{2}-1}\textrm{E}_{1-c_{s}^{2}}(2\sigma\tau^{\prime})]\right.\nonumber \\
 &  & \left.+\frac{a_{3}}{\tau_{0}}e^{\sigma\tau_{0}}[\tau_{0}\textrm{E}_{2-3c_{s}^{2}}(\sigma\tau_{0})-\tau\left(\frac{\tau_{0}}{\tau}\right)^{2-3c_{s}^{2}}\textrm{E}_{2-3c_{s}^{2}}(\sigma\tau)]\right\} .\label{eq:sol_02}
\end{eqnarray}
In the leading order, we see $E(\tau)\sim\frac{\tau_{0}}{\tau}x(\tau)\sim\frac{1}{\tau}e^{-\sigma\tau}$,
i.e. the electric field decays in the conducting medium \citep{Shokri:2017xxn}.
In the leading order, $y(\tau)\sim1$ means $n_{5}\sim\frac{\tau_{0}}{\tau}$.
We also see that when $c_{s}^{2}=1/3$, the analytic solutions of
$E(\tau)$ and $n_{5}(\tau)$ have the same form as in Eq. (\ref{eq:EN_A_01})
in previous subsection. 

In Figs. \ref{fig:ET}, \ref{fig:NT_1}, \ref{fig:endT_1}, we plot
the normalized $E/E_{0}$, $n_{5}/n_{5,0}$ and $\varepsilon/\varepsilon_{0}$
as functions of the proper time $\tau$. We choose the $\tau_{0}=0.6\textrm{ fm/c}$,
the speed of sound $c_{s}^{2}=1/3$ and $E_{0}^{2}/\varepsilon_{0}=0.1$.
The solid lines in those figures are the numerical results from Eqs.
(\ref{eq:xyz_01}), while the dashed lines are from approximate analytic
solutions (\ref{eq:sol_02}). We see that the approximation works
very well for small $a_{i}$.

In Fig. \ref{fig:ET}, we find $E/E_{0}$ is almost independent of
$a_{2}$ and $a_{3}$, as expected in Eq. (\ref{eq:sol_02}). The
$E/E_{0}$ decays rapidly as $a_{1}$ or $\sigma$ grows. Similar
to the cases in Subsec. \ref{subsec:For-EoS-1}, $E/E_{0}$ can be
negative at the late proper time. Such a behavior may come from the
competition between the anomalous conservation equation $\partial_{\mu}j_{5}^{\mu}=-CE\cdot B$
and Maxwell's equations.

In Fig. \ref{fig:NT_1}, the numerical results show that $n_{5}$
is almost independent of $a_{1}$ and $a_{3}$ in small $a_{i}$ cases
as expected in Eq. (\ref{eq:sol_02}). The $n_{5}$ decays slowly
as $a_{2}$ increases and the decay behavior of $n_{5}$ is also not
sensitive to variation of $\sigma$. 

In Fig. \ref{fig:endT_1}, we find that the time evolution of $\varepsilon(\tau)$
seems to be insensitive to $a_{1}$ and $a_{2}$. Because $E_{0}^{2}/\varepsilon_{0}\ll1$,
the contribution from the second term in Eq. (\ref{eq:sol_02}) which
is proportional to $\sigma E_{0}^{2}/\varepsilon_{0}$ is negligible.
Interestingly, the energy density decays slower as $a_{3}$ grows.
As shown in Fig. \ref{fig:endT_1}, for a large value of $a_{3}$,
e.g. $a_{3}=3.0$, the energy density even increases at early time.
That is because the fluid gain the energy from the EM fields, i.e.
the $a_{3}$ term in Eq. (\ref{eq:sol_02}) dominates. Similar behavior
is also found in the ideal MHD with a background magnetic field \citep{Pu:2016ayh,Roy:2015kma}.

We make some remarks here. From analytic solutions (\ref{eq:EN_A_01})
and (\ref{eq:sol_02}), we conclude that the CME and chiral anomaly
as quantum corrections play a role to the time evolution of the electric
field $E(\tau)$, the chiral charge density $n_{5}(\tau)$ and the
energy density $\varepsilon(\tau)$. With an initial magnetic field
parallel to the electric field (with $\chi=1$) and all $a_{i}$ ($i=1,2,3$)
are positive, $E(\tau)$/$n_{5}(\tau)$ decay faster/slower than the
cases without CME. If the initial magnetic field is anti-parallel
to the electric field (with $\chi=-1$) and all $a_{i}$ are negative,
$E(\tau)$/$n_{5}(\tau)$ decay slower/faster than the cases without
CME. This behavior is consistent with the anomalous conservation equation
$\partial_{\mu}j_{5}^{\mu}=-CE^{\mu}B_{\mu}=C\chi E(\tau)B(\tau)$
combined with Maxwell's equations. For example, if $\chi=+1$, we
have $\partial_{\tau}(n_{5}\tau)=C\tau\chi E(\tau)B(\tau)>0$, implying
that $n_{5}(\tau)$ decays slower than the case $C=0$. From Eq (\ref{eq:Maxwell_02a}),
we have $\partial_{\tau}[E\tau\exp(\sigma\tau)]=-\tau\chi\xi B(\tau)<0$,
i.e. $E(\tau)$ decays faster than the case $C=0$. Such a behavior
is due to that the chiral charge density is converted from the magnetic
helicity. For $\chi=-1,$ the magnetic helicity will be converted
from the chiral charge density so the behavior is opposite. The numerical
results in Figs. \ref{fig:EN_n}, \ref{fig:end_n}, \ref{fig:ET},
\ref{fig:NT_1} and \ref{fig:endT_1} are consistent with the above
observation.


\subsection{Discussions}

\label{subsec:Discussion-and-physical}In Subsec. \ref{subsec:For-EoS-1}
and \ref{subsec:For-EoS-2}, we have obtained the approximate analytic
solutions in two types of EoS. From Eq.(\ref{eq:B_03}), the proper
time behavior of the magnetic field seems to be the same as the case
without CME and finite conductivity, i.e. in an ideal MHD \citep{Pu:2016ayh,Roy:2015kma,Pu:2016bxy}.
It seems to be counter-intuitive and inconsistent with the Maxwell's
equations. Our explanation is as follows. The $E^{\mu}$ and $B^{\mu}$
defined in the four vector form of EM fields in Eq. (\ref{eq:EB_def01})
are the fields in the co-moving frame of the fluid. The $B(\tau)$
in Eq.(\ref{eq:B_03}) is the length of the magnetic field three vector
$\mathbf{B}$. To show the explicit contribution from CME and finite
conductivity to each component of $\mathbf{B}$, we will compute EM
fields three vector in the laboratory frame.

From Eq. (\ref{eq:EM_tensor_01}), we observe that the EM field strength
tensor $F^{\mu\nu}$ as well as the energy-momentum tensor $T^{\mu\nu}$
and fluid velocity $u^{\mu}$ is measured in the laboratory frame.
According to the standard definitions of EM fields through the field
strength tensor $F^{\mu\nu}$, i.e. 
\[
\mathbf{E}_{L}^{i}=F^{i0},\;\mathbf{B}_{L}^{i}=-\frac{1}{2}\epsilon^{ijk}F^{jk},
\]
we can get the EM fields in the lab frame
\begin{eqnarray}
\mathbf{E}_{L} & = & (\gamma v^{z}B(\tau),\;\chi\gamma E(\tau),\;0),\nonumber \\
\mathbf{B}_{L} & = & (-\gamma v^{z}\chi E(\tau),\;\gamma B(\tau),\;0),\label{eq:EB_lab}
\end{eqnarray}
where in this subsection, we will use the lower index $L$ for the
EM fields in the laboratory frame and $E(\tau)$ and $B(\tau)$ are
the functions solved in previous Subsec. \ref{sec:Approximate-analytic-solutions}.
We find that in the lab frame $\mathbf{B}_{L}^{x}$ and $\mathbf{E}_{L}^{y}$
depend on the finite conductivity $\sigma$ and CME coefficient $\xi$
through $E(\tau)$.

Next, we will check the self-consistence of Maxwell's equations. We
will prove that the CME and finite conducting current will not generate
the EM fields in the $z$ direction, i.e. $\mathbf{E}_{L}^{z}$ and
$\mathbf{B}_{L}^{z}$ are always vanishing. From 
\begin{equation}
\nabla\times\mathbf{E}_{L}=-\partial_{t}\mathbf{B}_{L},
\end{equation}
we observe that with Eq. (\ref{eq:EB_lab}) the $\partial_{t}\mathbf{B}_{L}^{z}=0$
and $\partial_{y}\mathbf{E}_{L}^{z}=-\partial_{t}\mathbf{B}_{L}^{x}+\partial_{z}\mathbf{E}_{L}^{y}=0$
are automatically satisfied. With the solution (\ref{eq:B_03}), we
can also obtain that $\partial_{x}\mathbf{E}_{L}^{z}=\partial_{t}\mathbf{B}_{L}^{y}+\partial_{z}\mathbf{E}_{L}^{x}=0$.

Similarly from
\begin{equation}
\nabla\cdot\mathbf{E}_{L}=n_{e},\;\nabla\cdot\mathbf{B}_{L}=0,
\end{equation}
and Eq.(\ref{eq:EB_lab}), we can also obtain that $\partial_{z}\mathbf{E}_{L}^{z}=-\partial_{x}\mathbf{E}_{L}^{x}-\partial_{y}\mathbf{E}_{L}^{y}=0$
with $n_{e}=0$, and $\partial_{z}\mathbf{B}_{L}^{z}=-\partial_{x}\mathbf{B}_{L}^{x}-\partial_{y}\mathbf{B}_{L}^{y}=0$.

We will focus on the last equation
\begin{equation}
\nabla\times\mathbf{B}_{L}=\mathbf{j}_{e}+\partial_{t}\mathbf{E}_{L}.
\end{equation}
Different with the charge current in a static conductor, the charge
current $\mathbf{j}_{e}$ of a relativistic fluid includes two parts.
The part parallel to the fluid velocity $u^{\mu}$ read
\begin{equation}
\mathbf{j}_{e,\parallel}=\sigma\mathbf{E}_{L,\parallel}+\xi\mathbf{B}_{L,\parallel},\label{eq:j_par_01}
\end{equation}
and the other part perpendicular to the fluid velocity is given by
\begin{equation}
\mathbf{j}_{e,\perp}=\sigma\gamma(\mathbf{E}_{L}+\mathbf{v}\times\mathbf{B}_{L})_{\perp}+\xi\gamma(\mathbf{B}_{L}-\mathbf{v}\times\mathbf{E}_{L})_{\perp},\label{eq:j_pend_01}
\end{equation}
with $\mathbf{v}$ being the three vector of fluid velocity, i.e,
$u^{\mu}=\gamma(1,\mathbf{v})$. In our case, since the fluid moves
alone the $z$ direction, the charge current is given by
\begin{equation}
\mathbf{j}_{e}=\left[\gamma(\mathbf{E}_{L}^{y}+v^{z}\mathbf{B}_{L}^{x})+\xi\gamma(\mathbf{B}_{L}^{y}-v^{z}\mathbf{E}_{L}^{x})\right]\mathbf{e}_{y}.
\end{equation}
With Eq. (\ref{eq:EB_lab}), we find that $\partial_{y}\mathbf{B}_{L}^{z}=\partial_{t}\mathbf{E}_{L}^{x}+\partial_{z}\mathbf{B}_{L}^{y}=0$
and $\partial_{t}\mathbf{E}_{L}^{z}=0$. The space derivative of magnetic
field in the $z$ direction is
\begin{equation}
\partial_{x}\mathbf{B}_{L}^{z}=\partial_{z}\mathbf{B}_{L}^{x}-\sigma\gamma(\mathbf{E}_{L}^{y}+v^{z}\mathbf{B}_{L}^{x})-\xi\gamma(\mathbf{B}_{L}^{y}-v^{z}\mathbf{E}_{L}^{x})-\partial_{t}\mathbf{E}_{L}^{y},\label{eq:dBz_01}
\end{equation}
where the left-handed-side of above equation equals to the right-handed-side
of Eq. (\ref{eq:Maxwell_02a}). Thus, inserting our solutions in Eqs.
(\ref{eq:EN_A_01}, \ref{eq:sol_02}) yields $\partial_{x}\mathbf{B}_{L}^{z}=0$.

Since both time and space derivatives of $\mathbf{E}_{L}^{z}$ and
$\mathbf{B}_{L}^{z}$ vanish and initial $\mathbf{E}_{L}^{z}$ or
$\mathbf{B}_{L}^{z}$ are chosen to be vanishing, we can conclude
that in our setup the CME and conducting current will not generate
EM fields in the $z$ direction in the lab frame. While only the space-time
derivatives of EM fields in the transverse direction, e.g.$\mathbf{B}_{L}^{x}$
and $\mathbf{E}_{L}^{y}$, are non-vanishing. This is quite different
with the case of a static media, in which the CME current can induce
a circular magnetic field \citep{Akamatsu2013}.

Thirdly, we will discuss the Bjorken fluid velocity. Usually, we can
consider the right-handed-side of Eq. (\ref{eq:EM_con_02}), as the
covariant form of Lorentz force acting on the fluid. In the lab frame,
we can rewrite it as
\begin{equation}
F^{\nu\lambda}j_{e\lambda}=(j_{e,0}\mathbf{E}_{L},\;\mathbf{j}_{e}\times\mathbf{B}_{L}).
\end{equation}
Since we have chosen the $\mu_{e}=0$, the electric field will not
accelerate the fluid, i.e. the zeroth component $j_{e,0}\mathbf{E}_{L}=0$.
The other component $\mathbf{j}_{e}\times\mathbf{B}_{L}$ is the Lorentz
force driving by the magnetic field, where $\mathbf{j}_{e}$ is given
by Eqs. (\ref{eq:j_par_01}, \ref{eq:j_pend_01}). In our case, the
EM fields with Lorentz force $\mathbf{j}_{e}\times\mathbf{B}_{L}$
is analogy to the so-called the force free fields (e.g. also see the
discussion in the classical electrodynamics \citep{Chandrasekhar285,Woltjer489}
and recent studies in Ref. \citep{Hong:2012,Xia:2016any}). Through
Eqs. (\ref{eq:Du_01}, \ref{eq:energy_density_03}), we have already
shown that the EM fields in our setup will not modify the fluid velocity.

At last, we will check the consistence of (anomalous) current conservation
equations. Since Eqs. (\ref{eq:EN_A_01}, \ref{eq:sol_02}) are the
solutions of anomalous current equation $\partial_{\mu}j_{5}^{\mu}=-CE\cdot B$,
the anomalous current equation should be satisfied. Because EM fields
are independent on $x,y$, the charge current conservation equation
reduces to $\partial_{\mu}j_{e}^{\mu}=\partial_{t}j_{e,0}+\nabla\cdot\mathbf{j}_{e}=\partial_{z}\mathbf{j}_{e,z}$,
with $\mathbf{j}_{e,z}=\mathbf{j}_{e,\parallel}=\sigma\mathbf{E}_{L,z}+\xi\mathbf{B}_{L,z}=0$.
We can conclude that the (anomalous) current conservation equations
are satisfied.

Before we end this section, we make some remarks here. We have computed
the EM fields in the lab frame and found our solutions satisfy the
Maxwell's equations. In our setup, the CME and electric conducting
current will not generate the EM fileds in $z$ direction in lab frame.
It is quite different with the case in a static media. We have also
shown the Lorentz force will not accelerate the fluid. At last, we
have checked the self-consistence of (anomalous) current conservation
equations.

\section{Summary and conclusions}

\label{sec:Summary-and-conclusion}We have solved MHD equations with
longitudinal boost invariance and transverse EM fields in the presence
of the CME and finite electric conductivity. The MHD equations involve
the energy-momentum, the electric charge and chiral charge (anomalous)
conservation equations coupled with Maxwell's equations. We consider
two types of EoS corresponding to the large chiral chemical potential
and the high temperature cases respectively. For further simplification,
we consider the electric charge neutral fluid and set the electric
charge density $n_{e}$ and its corresponding chemical potential $\mu_{e}$
vanish. 

We assume the Bjorken form of the fluid velocity in the longitudinal
direction. To keep the fluid velocity unchanged, we obtain the four-vector
form of the electric and magnetic field which are orthogonal to the
fluid velocity. To solve the MHD equations, we treat the terms with
the anomaly constant which is proportional to the Planck constant
$\hbar$ as perturbations. This is equivalent to an expansion in $\hbar$.
Then we apply the non-conserved charge method to obtain the approximate
analytic solutions. The comparison of the analytic solutions with
the exact numerical results shows good agreement. 

Finally we compute the EM field in three-vector form in the lab frame
and show the contributions from the electric conductivity and the
CME. According to Maxwell's equations, in our setup, the CME and electrically
conducting current only modify the EM fields in the transverse direction
in the lab frame. The electric and magnetic field in the z-direction
does not grow with time and space. The Lorentz force only changes
the time evolution of thermodynamic quantities and does not accelerate
the fluid.

Our results can provide a future test of complete numerical simulations
of the MHD with the CME. Since the polarization of chiral fermions
in the strong magnetic field is different from the ordinary magnetization
which is called chiral Barnett effect \citep{Fukushima:2018osn},
the current method can be applied to study the magnetization effect
in the future. 
\begin{acknowledgments}
S.P. would like to thank Masoud Shokri for helpful discussions. QW
is supported in part by the National Natural Science Foundation of
China (NSFC) under Grant No. 11535012 and No. 11890713, the 973 program
under Grant No. 2015CB856902, and the Key Research Program of the
Chinese Academy of Sciences under the Grant No. XDPB09. SP is supported
by One Thousand Talent Program for Young Scholars. IS is supported
by the Chinese Academy of Sciences and The World Academy of Sciences
(CAS-TWAS) Scholarship.
\end{acknowledgments}

\bibliographystyle{h-physrev}
\bibliography{MHD}

\end{document}